\documentclass[%
reprint,
superscriptaddress,
 amsmath,amssymb,
 aps,
prx]{revtex4-2}
\usepackage{tikz}
\usetikzlibrary{arrows.meta}
\usetikzlibrary{positioning}
\usetikzlibrary{intersections}
\usepackage{graphicx}%
\usepackage{dcolumn}%
\usepackage{bm}%
\usepackage{subcaption}
\usepackage{booktabs}  %
\usepackage{siunitx}
\usepackage{multirow}
\usepackage{adjustbox}
\usepackage{placeins}
\usepackage{enumitem}
\usepackage{float}
\makeatletter
\let\newfloat\newfloat@ltx
\makeatother
\usepackage{algorithm}
\usepackage{algpseudocode}

\makeatletter
\renewcommand{\fnum@algorithm}{\algorithmname~\thealgorithm}
\makeatother

\colorlet{figred}{red} %

\begin{document}

\title{Contrastive Regularization of Machine Learning Potentials}%

\author{Dimitrios Tzivrailis}
\affiliation{Université Paris-Saclay, CEA, List, F-91120 Palaiseau Cedex, France}
\affiliation{LPTMS, CNRS, Université Paris-Saclay, 91405 Orsay, France}

\author{Georgios Sotiropoulos}
\affiliation{Université Paris-Saclay, CEA, List, F-91120 Palaiseau Cedex, France}
\affiliation{LPTMS, CNRS, Université Paris-Saclay, 91405 Orsay, France}

\author{Alberto Rosso}
\affiliation{LPTMS, CNRS, Université Paris-Saclay, 91405 Orsay, France}

\author{Eiji Kawasaki}
\affiliation{Université Paris-Saclay, CEA, List, F-91120 Palaiseau Cedex, France}

\begin{abstract}

Machine learning models are increasingly trained as regressors and then deployed as generators, from language models to learned simulators. In this setting, accuracy on held-out data says little about the distribution a model produces when it is sampled. We study this problem where the correct distribution is fixed by first principles, in machine learning interatomic potentials (MLIPs) trained by minimizing the mean squared error (MSE) on density functional theory (DFT) energies and forces, then used to run molecular simulations. The MSE never constrains the Boltzmann distribution that the potential defines, so an MLIP can match held-out DFT data and still drift, when sampled, into spurious minima that corrupt its thermodynamics. Building on the variational principle of statistical physics, we introduce Contrastive Regularized MSE (CRMSE), a post-training step in which the potential acts as its own energy-based model and learns to reproduce the distribution it generates, without any new DFT data. On molecules from the MD17 benchmark, CRMSE restores physical sampling and recovers DFT free-energy profiles, while keeping energy and force errors within chemical accuracy. Our results suggest that a model meant to be sampled must be trained on the distribution it produces, not only for pointwise accuracy.
\end{abstract}

\maketitle
\section{Introduction}
\usetikzlibrary{decorations.markings}

\begin{figure*}[t]
    \centering

    \begin{subfigure}{0.32\textwidth}
        \pgfmathsetseed{42} %
        \centering
        \begin{tikzpicture}[scale=0.62]
            \draw [-{Stealth[length=3mm]}, thick] (0,0) -- (0,6) node[above] {$E$};
            \draw [-{Stealth[length=3mm]}, thick] (0,0) -- (9,0) node[right] {$\mathbf{x}$};

            \draw[very thick, blue!70!black, name path=curve]
            plot[smooth, tension=0.65] coordinates {
                (0.0,3)
                (1.0,1.3)
                (2.0,3.8)
                (3.2,1.8)
                (4.2,1.8)
                (5.2,3.5)
                (6.2,0.5)
                (7.2,2.8)
                (8.2,2.0)
                (9.0,3.0)
            };

            \draw[blue!80!figred, thick, dashed, line cap=round, line join=round]
            plot[smooth, tension=0.5] coordinates {
                (4, 1.70)
                (3.20, 1.82)
                (4.3, 2.15)
                (4.2, 2.4)
                (3.1,2.15)
                (4.7,3.1)
                (3.0,2.4)
                (3.0,2.8)
                (5.1,3.57)
                (5.6,3.4)
                (5.5,3.1)
                (7.15,2.8)
                (5.5,2.7)
                (7.0,2.4)
                (5.8,1.82)
                (6.25, 0.75)
                (6.30, 0.60)
            };

            \draw[blue!70!black,
                postaction={
                    decoration={
                        markings,
                        mark=between positions 0.28 and 0.545 step 0.013 with {
                            \pgfmathparse{int(random(0,1))}
                            \ifnum
                                \pgfmathresult=1
                                \draw[fill=gray!60, draw=black, line width=0.4pt] (0,0) circle (1.6pt);
                            \fi
                        }
                    },
                    decorate
                }
            ]

            plot[smooth, tension=0.65] coordinates {
                (0.0,3) (1.0,1.3) (2.0,3.8) (3.2,1.8) (4.2,1.8)
                (5.2,3.5) (6.2,0.5) (7.2,2.8) (8.2,2.0) (9.0,3.0)
            };
            plot[smooth, tension=0.65] coordinates {
                (0.0,3)
                (1.0,1.3)
                (2.0,3.8)
                (3.2,1.8)
                (4.2,1.8)
                (5.2,3.5)
                (6.2,0.5)
                (7.2,2.8)
                (8.2,2.0)
                (9.0,3.0)
            };

            \node[gray!70!black, font=\scriptsize, align=center] at (3.66,3.95)
                {training region};

            \draw[gray!90, dashed, thick, rounded corners] (2.0,1.45) rectangle (5.25,4.35);

        \end{tikzpicture}
    \end{subfigure}
    \hfill
    \begin{subfigure}{0.32\textwidth}
        \pgfmathsetseed{42} %
        \centering
        \begin{tikzpicture}[scale=0.62]
            \draw [-{Stealth[length=3mm]}, thick] (0,0) -- (0,6) node[above] {$E$};
            \draw [-{Stealth[length=3mm]}, thick] (0,0) -- (9,0) node[right] {$\mathbf{x}$};

            \draw[very thick, blue!80!white, opacity=0.4, name path=ghost]
            plot[smooth, tension=0.65] coordinates {
                (0.0,3)
                (1.0,1.3)
                (2.0,3.8)
                (3.2,1.8)
                (4.2,1.8)
                (5.2,3.5)
                (6.2,0.5)
                (7.2,2.8)
                (8.2,2.0)
                (9.0,3.0)
            };

            \draw[very thick, figred!80!black, opacity=0.4] plot [smooth, tension=0.7]
            coordinates {
                (0.0,3.3)
                (1.0,1.6)
                (1.8,4.3)
                (3.2,1.85)
                (4.2,1.85)
                (5.4,4.0)
                (6.2,0.9)
                (7.2,3.3)
                (8.2,2.3)
                (9.0,3.4)
            };

            \draw[very thick, figred!80!black] plot [smooth, tension=0.7]
            coordinates {
                (0.1,3.9)
                (1.0,2.3)
                (1.7,4.7)
                (3.2,1.9)
                (4.25,1.9)
                (5.6,4.9)
                (6.2,1.6)
                (7.2,3.9)
                (8.2,3.)
                (9.0,4.4)
            };

            \draw[gray!90, dashed, thick, rounded corners] (2.0,1.45) rectangle (5.25,4.35);

            \path[
                postaction={
                    decoration={
                        markings,
                        mark=between positions 0.28 and 0.545 step 0.013 with {
                            \pgfmathparse{int(random(0,1))}
                            \ifnum
                                \pgfmathresult=1
                                \draw[fill=gray!60, draw=black, line width=0.4pt] (0,0) circle (1.6pt);
                            \fi
                        }
                    },
                    decorate
                }
            ]

            plot[smooth, tension=0.65] coordinates {
                (0.0,3) (1.0,1.3) (2.0,3.8) (3.2,1.8) (4.2,1.8)
                (5.2,3.5) (6.2,0.5) (7.2,2.8) (8.2,2.0) (9.0,3.0)
            };
            plot[smooth, tension=0.65] coordinates {
                (0.0,3)
                (1.0,1.3)
                (2.0,3.8)
                (3.2,1.8)
                (4.2,1.8)
                (5.2,3.5)
                (6.2,0.5)
                (7.2,2.8)
                (8.2,2.0)
                (9.0,3.0)
            };

        \end{tikzpicture}
    \end{subfigure}
    \hfill
    \begin{subfigure}{0.32\textwidth}
        \pgfmathsetseed{42} %
        \centering
        \begin{tikzpicture}[scale=0.62]
            \draw [-{Stealth[length=3mm]}, thick] (0,0) -- (0,6) node[above] {$E$};
            \draw [-{Stealth[length=3mm]}, thick] (0,0) -- (9,0) node[right] {$\mathbf{x}$};

            \draw[very thick, blue!80!white, opacity=0.4, name path=ghost]
            plot[smooth, tension=0.7]
            coordinates {
                (0.0,3)
                (1.0,1.3)
                (2.0,3.8)
                (3.2,1.8)
                (4.2,1.8)
                (5.2,3.5)
                (6.2,0.5)
                (7.2,2.8)
                (8.2,2.0)
                (9.0,3.0)
            };

            \draw[gray!90, dashed, thick, rounded corners] (2.0,1.45) rectangle (5.25,4.35);

            \draw[green!50!black, thick, dashed, line cap=round, line join=round]
            plot[smooth, tension=0.5] coordinates {
                (4, 1.70)
                (3.20, 1.82)
                (4.3, 2.15)
                (4.2, 2.4)
                (3.1,2.15)
                (4.7,3.1)
                (3.0,2.4)
                (3.0,2.8)
                (4.9,3.57)
                (2.5,3.4)
                (5.1,4.0)
                (2.2,4.2)
                (5.3,4.7)
                (2.0,5.0)
                (5.2,4.5)
                (2.3,3.8)
                (4.3,2.6)
            };

            \draw[very thick, green!30!black] plot [smooth, tension=0.7]
            coordinates {
                (0.2,5.3)
                (0.7,3.9)
                (1.3,6.)
                (3.,2.1)
                (4.4,2.1)
                (5.8,5.7)
                (6.4,2.9)
                (7.2,5.3)
                (8.2,4.3)
                (9.0,5.1)
            };

            \path[
                postaction={
                    decoration={
                        markings,
                        mark=between positions 0.28 and 0.545 step 0.013 with {
                            \pgfmathparse{int(random(0,1))}
                            \ifnum
                                \pgfmathresult=1
                                \draw[fill=gray!60, draw=black, line width=0.4pt] (0,0) circle (1.6pt);
                            \fi
                        }
                    },
                    decorate
                }
            ]

            plot[smooth, tension=0.65] coordinates {
                (0.0,3) (1.0,1.3) (2.0,3.8) (3.2,1.8) (4.2,1.8)
                (5.2,3.5) (6.2,0.5) (7.2,2.8) (8.2,2.0) (9.0,3.0)
            };
            plot[smooth, tension=0.65] coordinates {
                (0.0,3)
                (1.0,1.3)
                (2.0,3.8)
                (3.2,1.8)
                (4.2,1.8)
                (5.2,3.5)
                (6.2,0.5)
                (7.2,2.8)
                (8.2,2.0)
                (9.0,3.0)
            };

        \end{tikzpicture}
    \end{subfigure}
    \hfill

    \caption{
        Schematic effect of CRMSE training on the learned energy landscape.
        \textbf{(a)} Standard MSE training reproduces the energy well in the data-rich region, but leaves the landscape poorly controlled away from the training set, where deep ghost minima may appear.
        \textbf{(b)} CRMSE does not substantially modify the data-constrained basin; instead, it raises the energy of atypical configurations that are visited by the sampler, thereby creating a protective wall.
        \textbf{(c)} The resulting landscape preserves the physical basin while increasing the barrier against escape, which stabilizes MALA sampling.
    }
    \label{fig:pcd_top_only}
\end{figure*}

\begin{figure*}[t]
    \centering
    \includegraphics[width=1.\linewidth]{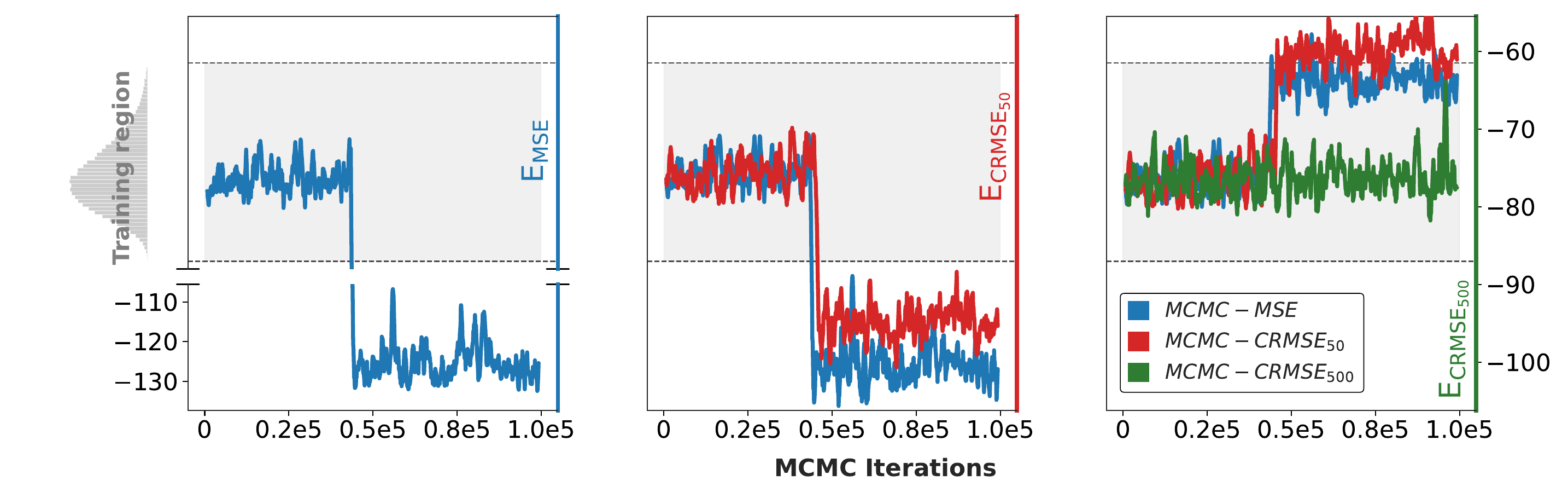}
    \caption{Energy of typical MALA chains as a function of MCMC iteration, evaluated under the MSE-trained potential ($E_{\rm MSE}$, left) and under CRMSE potentials after 50 and 500 post-training epochs ($E_{\rm CRMSE50}$, centre; $E_{\rm CRMSE500}$, right), all trained on 200 samples. The shaded band indicates the energy range explored by the DFT reference configurations. Under the MSE potential, the chain rapidly leaves this DFT energy window and becomes trapped in a spurious low-energy region. After 50 epochs of CRMSE post-training, the escape is delayed but not fully suppressed; after 500 epochs, the chain remains confined within the DFT energy window. The three panels therefore show the progressive stabilization of the sampler as contrastive post-training lifts the low-energy configurations visited by the escaping MSE chain. Full details of the post-training procedure and hyperparameters are given in Appendix~\ref{app:appendixE}.}
    \label{fig:combined_pcd_evolution}
\end{figure*}

Density Functional Theory (DFT)~\cite{PhysRev.136.B864,PhysRev.140.A1133}
provides a rigorous quantum-mechanical framework for
computing the electronic structure of atoms, molecules,
and materials, and has become an indispensable tool
across materials science, physics, chemistry, and
biology. Because DFT scales steeply with system size, it is
prohibitively expensive for the extended sampling required
to compute thermodynamic observables from first
principles, even for small molecules. This
bottleneck has motivated Machine Learning Interatomic
Potentials (MLIPs), data-driven surrogates that
approximate the DFT energy surface at a fraction of the
cost while retaining much of its
accuracy~\cite{PhysRevLett.98.146401, 10.1063/1.5019779, WANG2024109673}.
These models are most often Graph Neural Networks
(GNNs), whose architecture encodes the rotational and
permutational symmetries of atomic
systems~\cite{10.1063/1.5019779, pmlr-v198-wang22d}.
The dominant paradigm trains them by minimizing a Mean
Squared Error (MSE) loss on DFT-generated energies and
atomic forces, sometimes with adaptive weighting of the
two contributions~\cite{Ocampo2024Adaptive}. Evaluated on
held-out configurations, well-trained MLIPs reach
chemical accuracy on energies and forces.\footnote{Chemical accuracy is the usual quantum-chemistry target of 1~kcal/mol ($\approx$43~meV) between a computed and a reference energy, introduced by Pople~\cite{Pople1999Nobel} as the accuracy needed for reliable thermochemistry. There is no equivalent standard for forces, so, as is common in the MLIP literature, we use 1~kcal/mol/\AA{} as an informal benchmark.}

Yet predicting energies and forces is not why we build
these models. An MLIP is typically built to drive
molecular dynamics or Monte Carlo simulations, and what we
read off those
simulations are physical observables, such as free-energy profiles, structural distributions and conformational populations, each of them an average over the
equilibrium distribution that the potential itself
defines. A model can therefore be accurate at every
training point and still generate a meaningless
simulation, because nothing in its pointwise accuracy
controls the distribution it produces when sampled. This
is the simple idea that runs through the present work: a
potential meant to be sampled should be trained the way it is used: to reproduce its own sampling distribution, not merely to fit a finite set of data points. The lesson is not special to chemistry. MLIPs are one instance of a broader setting in AI: learned surrogates are often trained as regressors but then deployed as generators, through simulation or sampling. In this setting, the relevant object is not only the pointwise prediction of the energy given a configuration, but the probability distribution induced by repeatedly querying and sampling the model. Two examples make this concrete. Language models trained to predict the next token by maximum likelihood, a pointwise objective much like our MSE, can reach excellent held-out likelihood and still produce incoherent text when sampled directly, because errors in the model's low-probability tail are exposed and add up during generation~\cite{holtzman2020curious}. In model-based reinforcement learning, a dynamics model fitted to one-step transitions can be accurate for a single step, but when it is rolled out over many steps its errors compound, driving it into states it never saw during training~\cite{janner2019trust}. An MLIP is in the same situation. Its accuracy is checked once, on a fixed dataset, but during sampling its own forces decide where it goes next, so it reaches regions this check never covered.

Standard training misses this because the MSE is chasing
the wrong target. It is maximum-likelihood estimation
under the unphysical premise that energies are noisy measurements, which they are not, since the energy of a fixed configuration is deterministic. It also never refers to temperature or entropy, so nothing in it constrains the
distribution the model will be sampled from. The
consequence is concrete and unavoidable. The MSE pins the
energy only where data exist and leaves the rest of
configuration space free, and that rest is almost
everything: the fraction covered by a finite training set
shrinks exponentially with the number of atoms. In this
vast empty region an expressive network inevitably carves
spurious low-energy minima~\cite{unke_machine_2021},~\cite{Morrow2023Validate}, with
no penalty on held-out accuracy. A simulation that
explores long enough finds these holes, falls in, and
returns unphysical configurations and incorrect thermodynamics, that is, large errors in observables even when energies and forces are
accurate to chemical precision. The failure is statistical
rather than architectural, and it cannot be patched by a
better network.

The cure follows directly from the guiding idea: let the
model expose its own failure and correct it. We run short
simulations from the current potential, collect the
unphysical configurations it drifts into, and raise their
energy so the spurious basins are filled back in. We call
this Contrastive Regularized MSE (CRMSE): an MSE term keeps
the potential accurate on the training data, while a
contrastive term plays those data off against the model's
own simulated configurations, lowering the energy of the
data and raising it on the model samples. No new ab initio calculations are needed, since the model labels its own mistakes, and the whole thing runs as a cheap post-training step on top of an
already-trained MLIP.

This intuitive recipe is exactly what statistical physics
prescribes for matching a distribution. Reproducing the
Boltzmann distribution means minimizing a free energy,
equivalently a Kullback--Leibler divergence, which is the variational principle behind mean-field theory and the
Bogoliubov inequality~\cite{wu_solving_2019, Huembeli2022The}.
For an expressive potential this objective is intractable,
but it has a well-known tractable counterpart: the
contrastive training of energy-based
models~\cite{Du2019ImplicitGA, gagnon_clarifying_2022, Tieleman2008Training, Yan2021GraphEBM:},
which lowers the energy of configurations drawn from the
data and raises it on configurations drawn from the model, tracing back to contrastive divergence~\cite{Hinton2002CD}.
CRMSE adds this contrastive term to the MSE: the MSE keeps
the energies physical where data exist, and the contrastive
term reshapes the landscape everywhere else.

Boltzmann generators and normalizing-flow samplers train a separate generative model to reproduce a target distribution~\cite{Noe2019Boltzmann}; relative-entropy and force-matching coarse-graining fit a coarse-grained model to a reference all-atom ensemble~\cite{shell_relative_2008, noid_multiscale_2008}; and differentiable trajectory reweighting fits a model to measured observables~\cite{Thaler2021DiffTRe}. CRMSE is closest to relative-entropy minimization, but applied to the all-atom MLIP itself: the same network that supplies energies and forces is trained to reproduce the Boltzmann distribution it defines, using the training configurations as the reference ensemble.

That MLIPs grow unreliable when pushed to sample far from
their training data is well
documented~\cite{focassio_performance_2024, unke_machine_2021},
and the established remedies all inject information or bias
from outside the model: active learning labels the
offending configurations with fresh DFT calculations and
retrains~\cite{vandermause_--fly_2020}, and can be made more efficient by first driving the atomic configurations toward regions where the model is uncertain yet thermodynamically plausible, to decide which ones to label~\cite{SchwalbeKoda2021Adversarial}; committee or
ensemble uncertainty estimates flag them for labeling~\cite{Schran2020Committee}; and hand-built
repulsive priors forbid the worst of them a
priori~\cite{Yan2025Improving}; pretraining on a classical force field before fine-tuning on DFT data does the same in a softer way~\cite{YuanHeadGordon2025Teachers}. As we show in Sec.~\ref{sec:results_prior_reweighting}, a static geometric prior (harmonic walls on bond lengths and angles) is not enough on its own to restore correct sampling.
CRMSE injects nothing from outside. It acts on the same configurations as
active learning but scores them with the model alone,
adding no ab initio data, so the two are complementary
rather than equivalent. It also does less by design: it
restores the correct support of the sampling distribution,
confining the sampler to the physical basin and suppressing
the spurious minima it would otherwise fall into, but it
does not improve the potential's accuracy in those
off-manifold regions, which would demand data it never
sees.
Other groups have also tackled these instabilities without new ab initio data. Stability-Aware Boltzmann Estimator (StABlE) training~\cite{Raja2025StABlE} runs many simulations in parallel to find where an existing MLIP becomes unstable, and then corrects it using a reference observable such as a radial distribution function. Test-time adaptation~\cite{Cui2025TAIP} instead updates the potential during the simulation, through a self-supervised objective on the configurations it meets. CRMSE has the same goal, but unlike StABlE it needs no reference observable: the training configurations themselves are its only supervision, through the gradient of Sec.~\ref{sec:method}. And unlike test-time adaptation, it leaves the potential fixed once trained, so production simulations run on a single, unchanging energy function.

We demonstrate the method on the ethanol and aspirin
molecules of the MD17 dataset~\cite{Chmiela2017}, in its revised version rMD17~\cite{Christensen2020rMD17,Christensen2020}, using the GNN with local frame (GNN-LF) architecture as the surrogate potential~\cite{pmlr-v198-wang22d}. These are gas-phase systems of nine and twenty-one atoms, and MD17 is one of the most widely used benchmarks for machine-learned potentials.
We adopt this standard setup deliberately, with no new
reference data and no architectural changes, so that any
improvement is attributable to the training objective
alone. Standard MSE
training yields models that are accurate by every
regression metric, yet whose sampler drifts out of the
physical basin until the energies it visits lie up to about 10~kcal/mol below the lowest reference energies, with interatomic distances
stretched to unphysical values. The spurious minima
responsible are a statistical certainty already for a
molecule this small, not an artifact of large system size.
CRMSE keeps held-out force and energy errors within chemical accuracy while
returning the sampled configurations to the physical basin.
The stringent test is the free-energy profile along a
dihedral angle: it probes the relative populations
\emph{within} the basin, the part of the distribution that
confining the sampler would not by itself put right. CRMSE
recovers this profile to near-quantitative agreement with
the DFT reference, and does so even when the training set
is sharply reduced.

\section{Method}
\label{sec:method}

\subsection{Contrastive Regularization}

The central challenge in constructing a machine learning potential is to find
parameters $\theta$ such that $E_\theta(x)$ faithfully reproduces the ab initio
energy surface $E(x)$, where $x \in \mathbb{R}^{3n}$ denotes the vector of
Cartesian coordinates of a system of $n$ atoms,
\begin{equation}
    E_\theta(x) \approx E(x).
\end{equation}
In the MLIP literature, this is typically achieved by minimizing a Mean Squared Error (MSE) loss,
\begin{equation}
    \mathcal{L}^{\text{MSE}}(\theta) = \frac{1}{N} \sum_{i=1}^{N}
    \left( E_\theta(x_{i}) - E(x_{i}) \right)^{2},
\end{equation}
which constrains the values of $E_\theta(x_i)$ solely through the training dataset $\mathcal{D}$, consisting of configurations $x$ and their associated energies $E(x)$:
\begin{equation}
    \mathcal{D} = \{(x_i, E(x_i))\}_{i=1}^{N}
    \quad \text{with} \quad x_i \sim p(x),
\end{equation}
where $p(x) = e^{-\beta E(x)}/Z$ and $\beta = 1/(k_B T)$. In
practice, $\mathcal{L}^{\text{MSE}}$ is augmented by the
standard force-matching term, which penalizes the
discrepancy between the predicted forces $F_\theta(x) =
-\nabla_x E_\theta(x)$ and the reference forces $F(x)$. We
keep the energy-only form throughout the derivation and
notation below for clarity, and use this force-augmented
objective in all numerical experiments of
Sec.~\ref{sec:results}.

Because the MSE loss provides no signal outside the training points $\{x_i\}$, the model is free to take arbitrarily low values throughout the regions of configuration space not covered by $\mathcal{D}$. This is not a deficiency of any particular architecture of $E_\theta$, but a fundamental limitation of regression on finite data: any sufficiently expressive model can develop spurious minima in unvisited regions without any degradation of accuracy on $\mathcal{D}$. The severity of this problem scales with the dimensionality of the system: the fraction of $\mathbb{R}^{3n}$ sampled by a finite training set vanishes exponentially with $n$, making spurious minima a statistical inevitability rather than a pathological edge case. In the MD setting this is particularly consequential: the simulation continuously generates new configurations by integrating the equations of motion, and any spurious minimum encountered will trap the trajectory, producing the unphysical configurations illustrated in Fig.~\ref{fig:pcd_top_only}(a).

To address this, we argue that the training objective must account for the full probability density implicitly defined by the learned potential,
\begin{equation}
    p_\theta(x) = \frac{e^{-\beta E_\theta(x)}}{Z_\theta}.
\end{equation}

The normalization of the Boltzmann distribution provides the necessary regularization: any configuration not well-represented in $\mathcal{D}$ should have low probability density under $p_\theta$, and hence high energy under $E_\theta$. In other words, the energies of configurations corresponding to spurious minima of $E_\theta(x)$ must be raised. We therefore
propose a contrastive regularization term $\mathcal{L}^{\text{CR}}(\theta)$ that directly optimizes $p_\theta(x)$ to match the target distribution $p(x)$, and combine it with the MSE loss into the \textit{Contrastive Regularized MSE} (CRMSE) objective. This is close to relative-entropy and contrastive coarse-graining~\cite{shell_relative_2008,Ding2022ContrastiveCG}, applied here to the all-atom MLIP itself.

The regularization term is derived by minimizing the Kullback--Leibler
divergence from the target distribution $p$ to the model distribution
$p_\theta$,
\begin{equation}
    D_{\mathrm{KL}}(p \| p_\theta)
    = \langle \log p(x) \rangle_p - \langle \log p_\theta(x) \rangle_p
    \geq 0.
\end{equation}
Since the first term does not depend on $\theta$, minimizing
$D_{\mathrm{KL}}(p \| p_\theta)$ with respect to $\theta$ is equivalent
to minimizing
\begin{equation}
    \langle -\log p_\theta(x) \rangle_p
    = \beta \langle E_\theta(x) \rangle_p + \log Z_\theta.
\end{equation}
This is the
negative log-likelihood of the training data under $p_\theta$,
and it plays the role of a variational free energy in the spirit
of the Bogoliubov inequality. Differentiating with respect to $\theta$ gives
\begin{equation}
    \begin{split}
        \nabla_\theta \langle -\log p_\theta(x) \rangle_p
        &= \beta \langle \nabla_\theta E_\theta(x) \rangle_p
        +  \nabla_\theta \log Z_\theta \\
        &= \beta \langle \nabla_\theta E_\theta(x) \rangle_p
        - \beta \langle \nabla_\theta E_\theta(x) \rangle_{p_\theta}.
    \end{split}
\end{equation}
The contrastive regularization term is therefore a maximum likelihood
objective, shaping $E_\theta$ so that observed configurations are
assigned high probability and all others low probability.

The two resulting terms are referred to respectively as the positive term and negative term contributions. The positive term lowers the energies of training configurations $x_i \sim p(x)$, i.e., those present in the training set. The negative term arises from the partition function $Z_\theta$ (i.e. the normalization constant ensuring that the Boltzmann distribution integrates to unity) and raises the energies of configurations under $p_\theta(x)$, thereby suppressing their probability density. The method is termed \textit{contrastive} because the loss is defined by the opposition between these two populations: the training set samples $x \sim p(x)$, whose energies are pushed down, and the model samples $x \sim p_\theta(x)$, whose energies are pushed up. This is the same push-pull structure as in contrastive divergence~\cite{Hinton2002CD}; the persistent version we use (Sec.~\ref{sec:selfconsistent}) is Tieleman's PCD~\cite{Tieleman2008Training}.

In practice, we approximate the contrastive regularization loss term via Monte Carlo sampling:
\begin{equation}
    \mathcal{L}^{\text{CR}}(\theta)
    = \frac{1}{N} \sum_{i=1}^{N} E_\theta(x_i)
    - \frac{1}{M} \sum_{j=1}^{M} E_\theta(x_j^-),
\end{equation}
where $x_i$ denotes the "positive" configurations drawn from the training set samples, and $x_j^-$ are the "negative" configurations obtained by sampling the learned distribution $p_\theta(x)$. The gradient of this loss is that of the Kullback--Leibler divergence derived above, up to the constant factor $\beta$, which we absorb into the regularization weight $\lambda$ below; all values of $\lambda$ reported in this work follow this convention.

The full training objective, the CRMSE loss, is
\begin{equation}
    \mathcal{L}^{\text{CRMSE}}(\theta) = \mathcal{L}^{\text{MSE}}(\theta)
    + \lambda\, \mathcal{L}^{\text{CR}}(\theta).
    \label{eq:crmse}
\end{equation}
The hyperparameter $\lambda > 0$ controls the relative weight of the regularization term. In our experiments we find that values in the range $\lambda \in [0.01, 0.5]$  yield stable training across the systems studied; a systematic sensitivity analysis is provided in Appendix~\ref{app:lambda}.

Importantly, CRMSE is designed as a post-training procedure: the initial parameters $\theta_0$ are taken from an MLIP already trained by standard MSE minimization, and the contrastive regularization is applied on top without modifying the underlying architecture or requiring additional ab initio data. Practitioners can therefore correct an existing, potentially unreliable MLIP at a fraction of the cost of full retraining, making CRMSE applicable to models produced by any standard pipeline.

\subsection{Self-Consistent Post-Training}
\label{sec:selfconsistent}

Evaluating $\mathcal{L}^{\text{CR}}(\theta)$ requires drawing negative samples $x_j^- \sim p_\theta(x)$, which is non-trivial since $Z_\theta$ is intractable. We use the Unadjusted Langevin Algorithm (ULA)~\cite{Roberts1996Langevin}, which approximates samples from $p_\theta$ by iterating
\begin{equation}
    x_j^{-,t+1} = x_j^{-,t} - \eta\,\nabla_{x} E_\theta(x_j^{-,t})
    + \sqrt{2\eta/\beta}\;\xi^t,
    \quad \xi^t \sim \mathcal{N}(0, \mathbf{I}_{3n}),
    \label{eq:ula}
\end{equation}
where $\eta$ is the step size and $\mathbf{I}_{3n}$ is the $3n$-dimensional identity matrix. Each ULA step moves the configuration toward lower-energy regions of $E_\theta$ while the noise term, scaled by the temperature $\beta^{-1}$, ensures broad exploration of $p_\theta(x)$. In practice we use the per-atom preconditioning of Appendix~\ref{app:precond}.

To avoid restarting chains from scratch at each gradient step, we adopt \textit{Persistent Contrastive Divergence} (PCD): the negative samples $\{x_j^-\}$ are stored in a replay buffer $\mathcal{B}$ and carried over across updates, with only $K$ additional ULA steps applied per iteration. Although $K$ steps cover only a limited distance in $\mathbb{R}^{3n}$ per update, the persistent buffer is refreshed continuously over many updates, so the choice of initialization is not critical to the final result; it affects only how quickly the negative chains populate the relevant regions. The post-training setting nonetheless offers a convenient warm start: the MSE-MALA chains that motivated CRMSE are already available and lie in the atypical, low-energy regions the contrastive term must penalize, so seeding $\mathcal{B}$ with them accelerates training and avoids a burn-in phase. Importantly, this does not require identifying the spurious minima explicitly, nor enumerating them all: even when the seed configurations cover only a subset of the spurious minima, the stochastic ULA exploration drives the negative chains to discover and suppress the remaining ones over the course of training.

As $\theta$ is subsequently updated, the buffer is continuously refreshed, creating a self-consistent feedback loop: the current $E_\theta$ drives the ULA chains, the refined chains supply the negative batch, and the resulting gradient update corrects $E_\theta$. Since all energy evaluations are performed under the learned potential $E_\theta$ rather than the ab initio reference, the total computational overhead is modest relative to the original MSE training. The full procedure is given in Algorithm~\ref{alg:crmse}.

\begin{algorithm}
    \caption{CRMSE Post-Training}
    \label{alg:crmse}
    \begin{algorithmic}

        \Require MSE-pretrained parameters $\theta_0$,
        training set $\mathcal{D} = \{(x_i, E(x_i))\}_{i=1}^{N}$,
        regularization weight $\lambda$,
        ULA step size $\eta$,
        ULA steps per update $K$

        \State Initialize $\theta \leftarrow \theta_0$
        \State Initialize replay buffer $\mathcal{B}$
        \quad \textit{(unphysical configs from MSE-MALA chains)}

        \Repeat
            \State Sample negative batch $\{x_j^-\} \sim \mathcal{B}$
            \quad \textit{(persistent chains)}
            \For{$k = 1$ to $K$}
                \quad \textit{(ULA step)}
                \State $x_j^- \leftarrow x_j^-
                - \eta\,\nabla_x E_\theta(x_j^-)
                + \sqrt{2\eta/\beta}\;\xi$,
                \quad $\xi \sim \mathcal{N}(0, \mathbf{I}_{3n})$
            \EndFor
            \State Update $\mathcal{B}$ with refined $\{x_j^-\}$
            \State $\mathcal{L}^{\text{MSE}} \leftarrow
            \dfrac{1}{N}\displaystyle\sum_i
            \left(E(x_i) - E_\theta(x_i)\right)^2$
            \State $\mathcal{L}^{\text{CR}} \leftarrow
            \dfrac{1}{N}\displaystyle\sum_i E_\theta(x_i)
            - \dfrac{1}{M}\displaystyle\sum_j E_\theta(x_j^-)$
            \State $\theta \leftarrow \theta
            - \alpha\,\nabla_\theta
            \bigl(\mathcal{L}^{\text{MSE}}
            + \lambda\,\mathcal{L}^{\text{CR}}\bigr)$
            \quad \textit{($\alpha$: learning rate)}
        \Until{convergence}
        \label{crmse-algorithm}
    \end{algorithmic}
\end{algorithm}

The effect of CRMSE post-training on the energy landscape is illustrated schematically in Fig.~\ref{fig:pcd_top_only}. Panel~(a) shows the landscape produced by standard MSE training: $E_\theta$ accurately reproduces the ab initio surface within the data-rich training region, but is poorly constrained elsewhere, allowing deep spurious minima to develop. Panel~(b) shows the effect of the contrastive post-training step: the negative ULA chains visit these atypical configurations, and the push-up term raises their energies without substantially distorting the data-constrained basin. This creates a protective energy wall around the physical region. Panel~(c) shows the resulting landscape: the spurious minima are suppressed, the physical basin is preserved, and the barrier against escape is increased, which stabilizes subsequent molecular dynamics sampling. The empirical counterpart of this schematic is shown in Fig.~\ref{fig:combined_pcd_evolution}, which reports actual MALA energy trajectories generated with the MSE-trained potential and with CRMSE potentials at successive stages of post-training. Under the MSE potential, the chain rapidly leaves the DFT energy window and becomes trapped in a spurious low-energy minimum. After partial CRMSE post-training this escape is delayed but not fully suppressed, whereas after full post-training the chain remains confined within the energy range explored by the DFT reference. This shows directly how CRMSE stabilizes sampling: as post-training proceeds, the spurious low-energy regions reached by the escaping MSE trajectory are progressively lifted, removing the downhill routes that leave the physical basin.

\section{Results}
\label{sec:results}

We now demonstrate the CRMSE post-training procedure on a molecular system generated by ab initio molecular dynamics. Our goal is twofold. First, we establish that a neural network potential trained by standard MSE minimization can reproduce held-out DFT energies and forces with high accuracy while nonetheless failing to generate physically meaningful samples, both with MALA and with Langevin dynamics. Second, we show that the contrastive regularization of Sec.~\ref{sec:method} corrects this failure without degrading the predictive accuracy of the original model.

We focus on the ethanol molecule from the MD17 dataset, a nine-atom system whose 27-dimensional configuration space reduces to 21 effective dimensions once global translations and rotations are removed.

\begin{figure}[h!]
    \centering
    \includegraphics[width=1.0\linewidth]{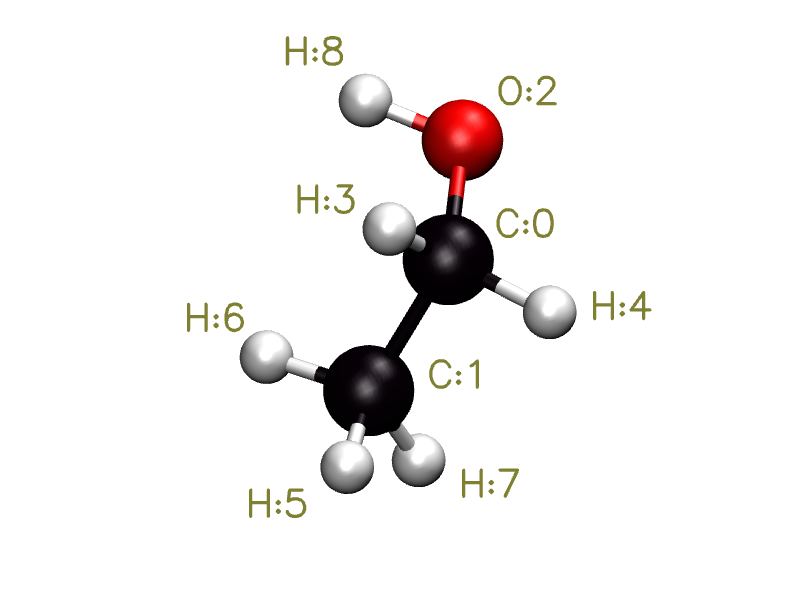}
    \caption{Representative ethanol molecule from the MD17 DFT dataset, with atoms labeled by type and index.}
    \label{fig:dft_eth_mol}
\end{figure}

A representative configuration is shown in Fig.~\ref{fig:dft_eth_mol}, with atoms labeled by type and index. The system is small enough to permit extensive sampling and direct comparison against the reference Boltzmann distribution, yet large enough that the spurious low-energy minima identified in Sec.~\ref{sec:method} arise generically rather than as a pathological edge case. Throughout, we use the GNN-LF architecture of Ref.~\cite{pmlr-v198-wang22d} as the surrogate potential $E_\theta$ and benchmark samples against the DFT reference using two observables: the sampled energy distribution and the distributions of selected interatomic distances. Although our conclusions apply equally to molecular dynamics and unadjusted Langevin (ULA) sampling, we sample with the Metropolis-adjusted Langevin algorithm (MALA)~\cite{rossky_brownian_1978},~\cite{Roberts1996Langevin}, whose accept/reject step permits the larger step sizes needed to explore the energy landscape efficiently, in contrast to the asymptotically vanishing steps required for ULA. We check this in Sec.~\ref{sec:ula_validation} with BAOAB, a Langevin integrator that keeps the inertia of the atoms and therefore produces an MD trajectory.

\subsection{Surrogate accuracy and the sampling failure}
\label{sec:results_failure}

\begin{figure}[h!]
\centering
\captionsetup{justification=raggedright, singlelinecheck=false, format=plain}
    \includegraphics[width=1.\linewidth]{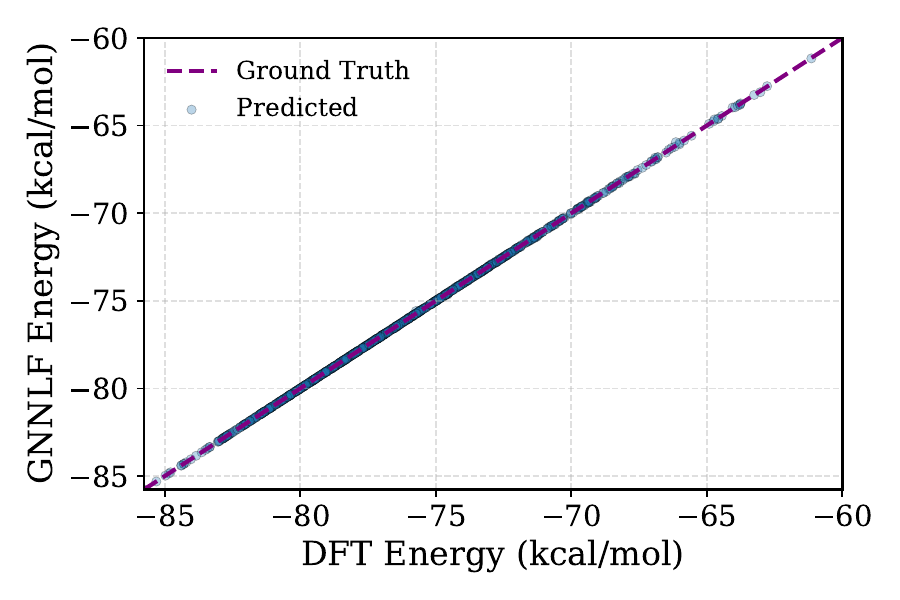}
    \includegraphics[width=1.\linewidth]{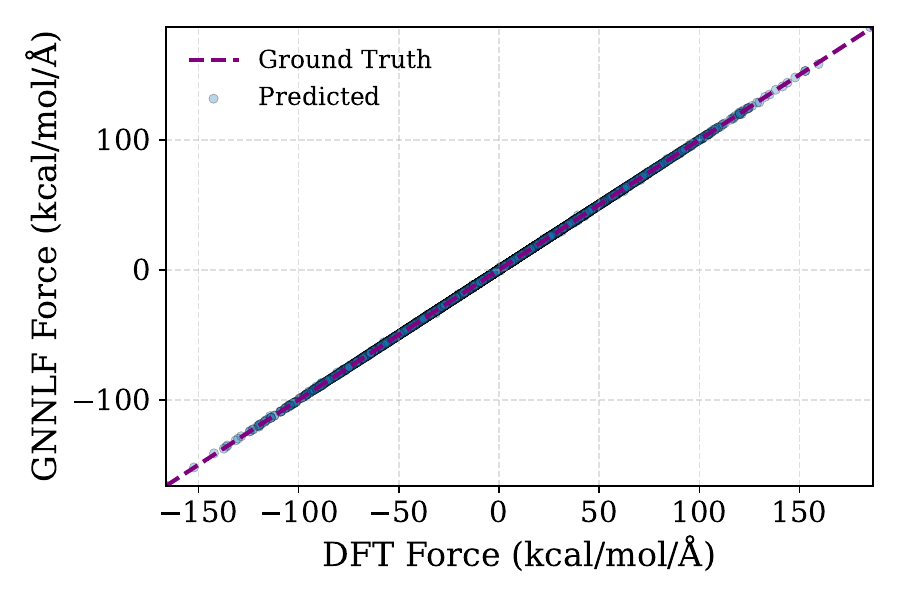}
    \caption{Performance of the MSE-trained GNN-LF potential on 1000 held-out DFT configurations: predicted versus reference energies (top) and forces (bottom).}
    \label{fig:mse_perf}
\end{figure}

We first verify that standard MSE training yields an accurate surrogate. We train the GNN-LF potential on 950 DFT-generated ethanol configurations, holding out 50 for validation, by minimizing the energy-and-force MSE objective with the Adam optimizer over 6000 epochs. We use 950 configurations because the rMD17 dataset authors explicitly recommend training on no more than 1000 structures: the configurations derive from a molecular-dynamics trajectory and are temporally correlated, so larger subsets add no statistically independent information~\cite{Christensen2020}. Our setup therefore uses the maximum number of effectively independent configurations the benchmark provides, and the sampling failure reported below cannot be dismissed as an artifact of an undersized training set. Evaluated on 1000 held-out DFT reference configurations, the model reproduces both energies and forces with high fidelity: as shown in Fig.~\ref{fig:mse_perf}, the predictions fall tightly along the reference line across the full range of the test set, with no systematic deviation. Quantitatively, the energy and force errors on the held-out set are reported in Table~\ref{tab:errors} and are at the level expected for a well-converged MLIP.

\begin{table}[h!]
    \centering
    \caption{Energy and force mean absolute errors (MAE) of the MSE-pretrained and CRMSE post-trained models on the 1000-configuration held-out DFT test set. CRMSE
introduces a small increase in both errors, but both remain well within
chemical accuracy, confirming that the contrastive regularization does not
meaningfully degrade predictive accuracy on the data-constrained region.}
    \label{tab:errors}
    \begin{ruledtabular}
    \begin{tabular}{lcc}
         & Energy MAE & Force MAE \\
         & (kcal/mol) & (kcal/mol/\AA) \\
        \colrule
        MSE   & 0.01 & 0.07 \\
        CRMSE & 0.03 & 0.13 \\
    \end{tabular}
    \end{ruledtabular}
\end{table}

\begin{figure}[ht!]
    \centering
    \includegraphics[width=1.\linewidth]{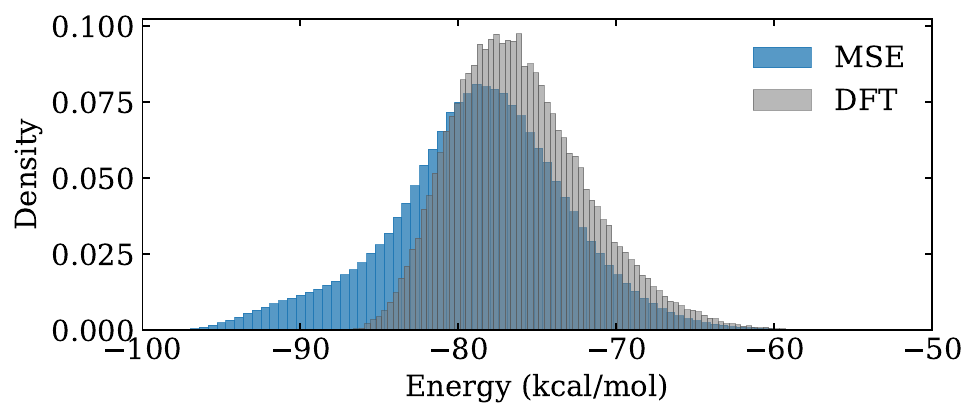}
    \caption{Sampled energy distribution for configurations sampled by MALA under the MSE-trained model (blue), compared with the DFT reference (gray). The horizontal axis is the GNN-predicted energy $E_\theta$ rather than the DFT energy: the sampled configurations lie outside the dataset, so their DFT energies are not available, and both distributions are placed on the model's common energy scale. Data are aggregated over 32 independent chains of $10^6$ steps each.}
    \label{fig:mse_dos}
\end{figure}

This pointwise accuracy, however, does not survive sampling. The MD17 trajectories were nominally generated at $500$~K, but their internal degrees of freedom correspond to a higher temperature: about $650$~K for ethanol and $555$--$580$~K for aspirin. The most likely reason is that the thermostat set the kinetic energy for all $3N$ degrees of freedom of the molecule, while only its $3N-6$ internal ones carry it (Appendix~\ref{app:effective_temperature}). We incorporate the trained potential into MALA at $k_BT = 1.3$~kcal/mol, using a global update with a per-atom preconditioning matrix (defined in Appendix~\ref{app:precond}). Initializing each chain from a reference DFT configuration, we run 32 independent simulations of $10^6$ steps with step size $0.0005$. Despite the model's predictive fidelity, the trajectories escape the physical basin and settle into spurious low-energy minima absent from the reference data. Figure~\ref{fig:mse_dos} shows that the resulting sampled energy distribution is shifted well below the DFT distribution, confirming that the sampler fails to reproduce the target Boltzmann distribution; the 32 chains are mutually consistent, so this shift reflects a systematic failure of the potential rather than chain-to-chain variance.

\begin{figure}[ht!]
    \centering
    \begin{subfigure}{\linewidth}
        \centering
        \includegraphics[width=\linewidth]{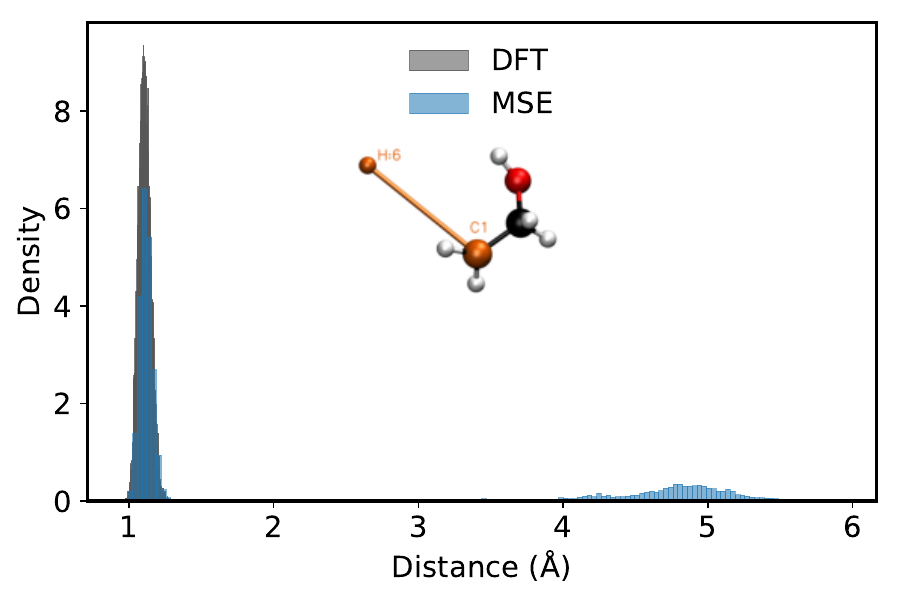}
        \caption{}
        \label{fig:mse_dist_a}
    \end{subfigure}

    \vspace{0.3cm}

    \begin{subfigure}{\linewidth}
        \centering
        \includegraphics[width=\linewidth]{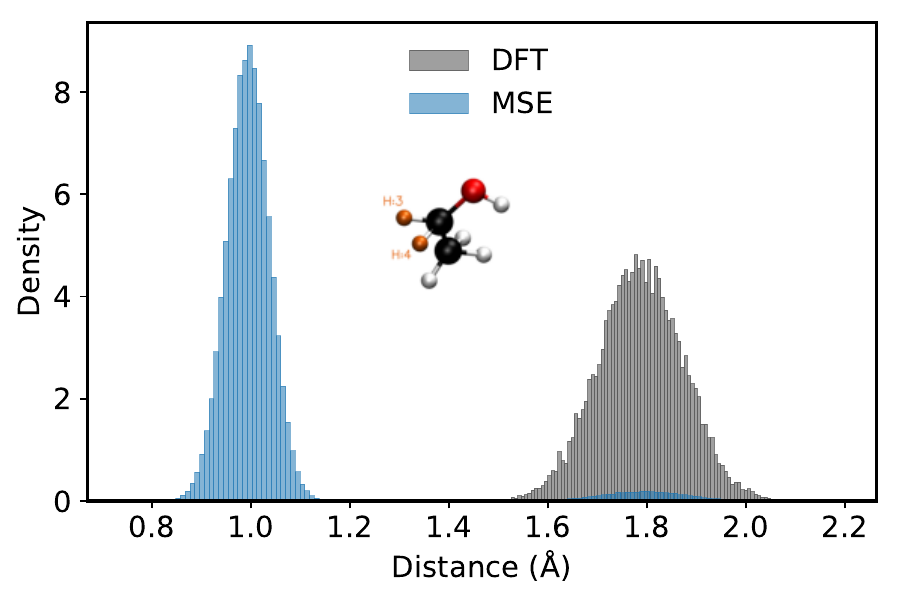}
        \caption{}
        \label{fig:mse_dist_b}
    \end{subfigure}

    \vspace{0.3cm}

    \begin{subfigure}{\linewidth}
        \centering
        \includegraphics[width=\linewidth]{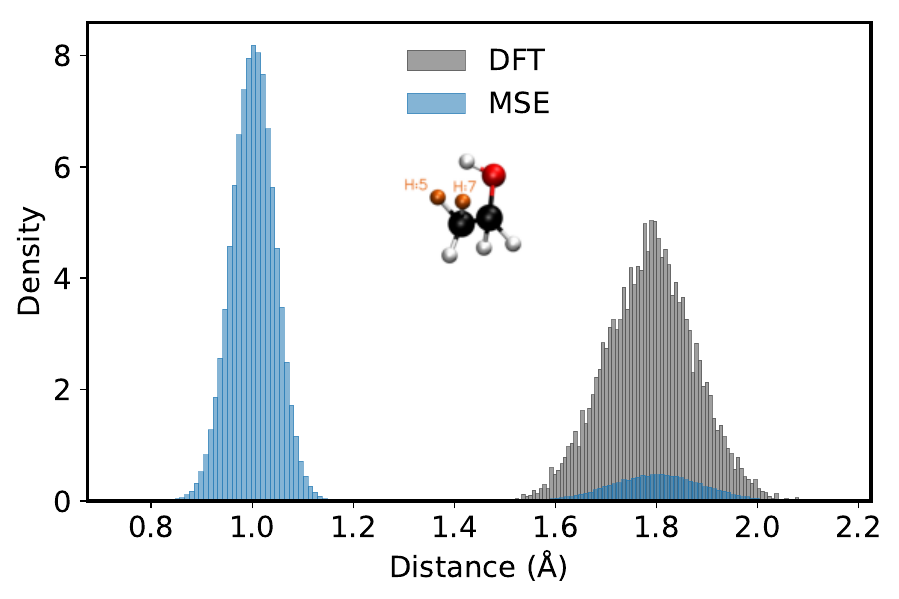}
        \caption{}
        \label{fig:mse_dist_c}
    \end{subfigure}

    \caption{Densities of interatomic distances for selected atom pairs under MSE-MALA sampling (blue) and the DFT reference (gray).}
    \label{fig:mse_dist}
\end{figure}

The failure is equally visible in the molecular geometry. Fig.~\ref{fig:mse_dist} reports the histograms of interatomic distances for selected atom pairs: the sampled configurations contain atoms stretched far beyond physical bond lengths and others driven unphysically close, with the distributions bearing no resemblance to the sharply peaked DFT reference. A model that is excellent by every regression metric is therefore unreliable as a generative sampler. This is the central failure mode that motivates contrastive regularization. Others have reported the same gap between pointwise accuracy and simulation quality. Fu et al.~\cite{Fu2023ForcesNotEnough} found that low force errors do not reliably predict MD stability or the accuracy of observables, across several architectures and including MD17 molecules. Stocker et al.~\cite{Stocker2022Robust} found that low test-set errors do not guarantee stable dynamics for organic molecules at elevated temperature, with pathologies sometimes appearing only after hundreds of picoseconds, and Morrow et al.~\cite{Morrow2023Validate} list false minima and maxima in potentials with low held-out error as a separate failure to look for when validating a potential. Our contribution is to explain where this failure comes from, namely the missing normalization in the MSE objective (Sec.~\ref{sec:method}), and to fix it without new reference data.

\subsection{CRMSE correction}
\label{sec:results_crmse}

We now apply the CRMSE post-training of Sec.~\ref{sec:method}, seeding the persistent replay buffer with configurations drawn from the failed MSE-MALA chain. As discussed in Sec.~\ref{sec:method}, this seeding is a convenient warm start rather than a requirement: it places the negative chains in the atypical regions the contrastive term must penalize and so accelerates training, but the spurious minima need not be identified explicitly. We construct the buffer by partitioning a $10^6$-step MSE-MALA chain into segments of $10^3$ steps and drawing one configuration from each, yielding 1000 persistent particles that span the full range of distances from the reference manifold, from near-equilibrium to strongly unphysical. This buffer is drawn from a single one of the 32 MALA chains and therefore does not contain every spurious minimum the sampler can reach; consistent with the discussion of Sec.~\ref{sec:method}, the negative chains nonetheless discover and suppress the remaining minima during post-training, so that the corrected model confines all 32 production chains to the physical basin.

Starting from the MSE-pretrained weights and restoring the Adam optimizer state (moments and learning rate, ${\approx}10^{-5}$) from the last epoch of the MSE training, we train for 200 epochs with a batch size of 16 and regularization weight $\lambda = 0.015$. At each epoch the persistent particles evolve for $K = 200$ ULA steps with step size $0.0004$, smaller than the MALA sampling step size to keep the negative-chain propagation stable under the still-evolving potential $E_\theta$. The standard route to repairing an unreliable potential is active learning, which labels the offending configurations with additional DFT single points and appends them to the training set before retraining. This route is not directly available on a fixed public benchmark such as MD17: we did not generate the reference trajectories, and producing new labels consistent with the original level of theory falls outside a benchmark comparison. We therefore take the dataset as published and measure CRMSE against the same MSE-trained potential it post-trains, using no new ab initio information; we return to the relationship between CRMSE and active learning, and to what this minimal setup implies, in Sec.~\ref{sec:conclusion}.

\begin{figure}[ht!]
    \centering
    \captionsetup{justification=raggedright, singlelinecheck=false, format=plain}
    \includegraphics[width=1.\linewidth]{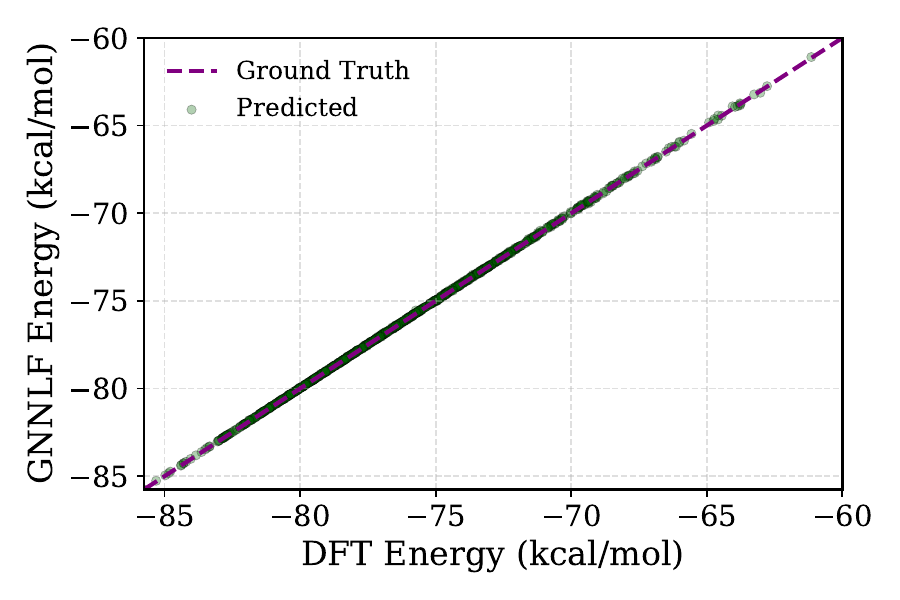}
    \includegraphics[width=1.\linewidth]{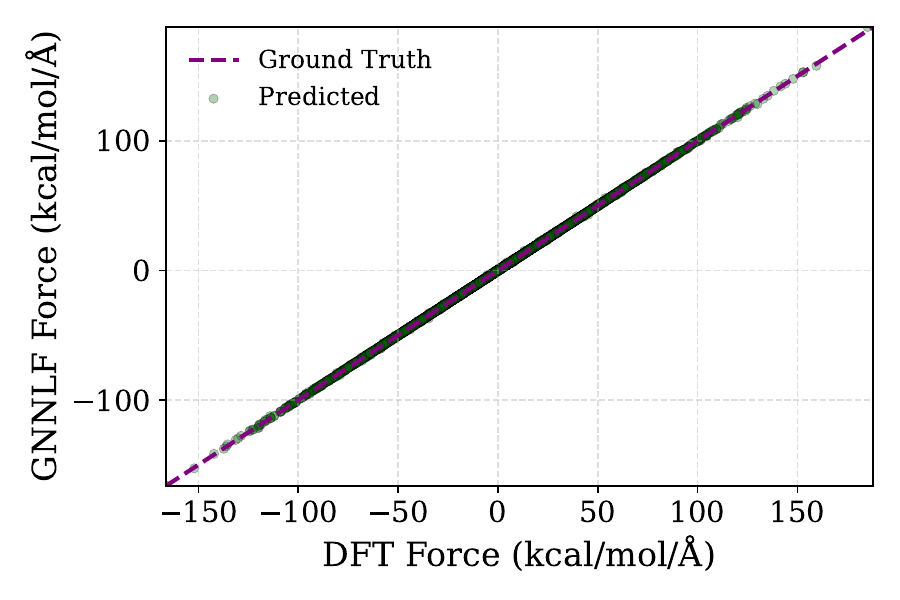}
    \caption{Performance of the CRMSE post-trained model on the held-out DFT test set: predicted versus reference energies (top) and forces (bottom). The accuracy closely matches that of the MSE-pretrained model (Fig.~\ref{fig:mse_perf}).}
    \label{fig:crmse_perf}
\end{figure}

After post-training, the held-out energy and force predictions remain close to those of  the MSE model (Fig.~\ref{fig:crmse_perf} and Table~\ref{tab:errors}), confirming that the regularization reshapes the unconstrained regions of the landscape without perturbing the data-constrained basin.

\begin{figure}[ht!]
    \centering
    \includegraphics[width=1.0\linewidth]{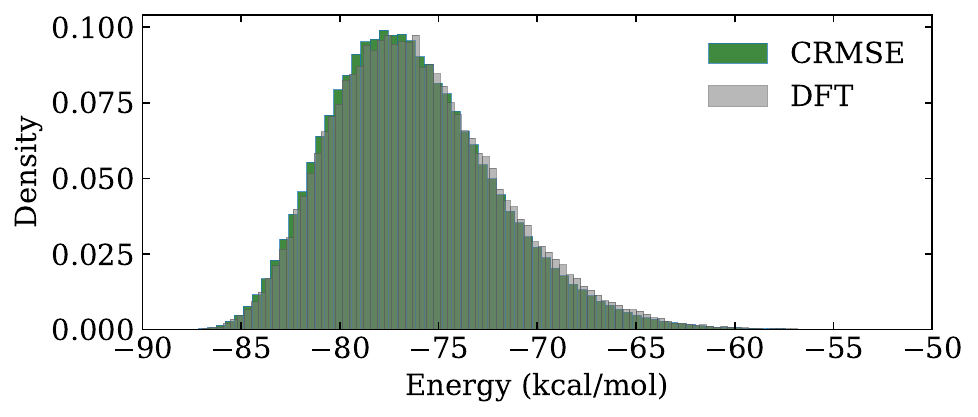}
    \caption{Energy distribution for configurations sampled by MALA under the CRMSE post-trained model (green), compared with the DFT reference (gray). As in Fig.~\ref{fig:mse_dos}, the horizontal axis is the GNN-predicted energy $E_\theta$ rather than the DFT energy, since the sampled configurations lie outside the dataset and their DFT energies are not available. Data are aggregated over 32 independent chains of $10^6$ steps each.}
    \label{fig:crmse_dos}
\end{figure}

\begin{figure}[ht!]
    \centering
    \begin{subfigure}{\linewidth}
        \centering
        \includegraphics[width=\linewidth]{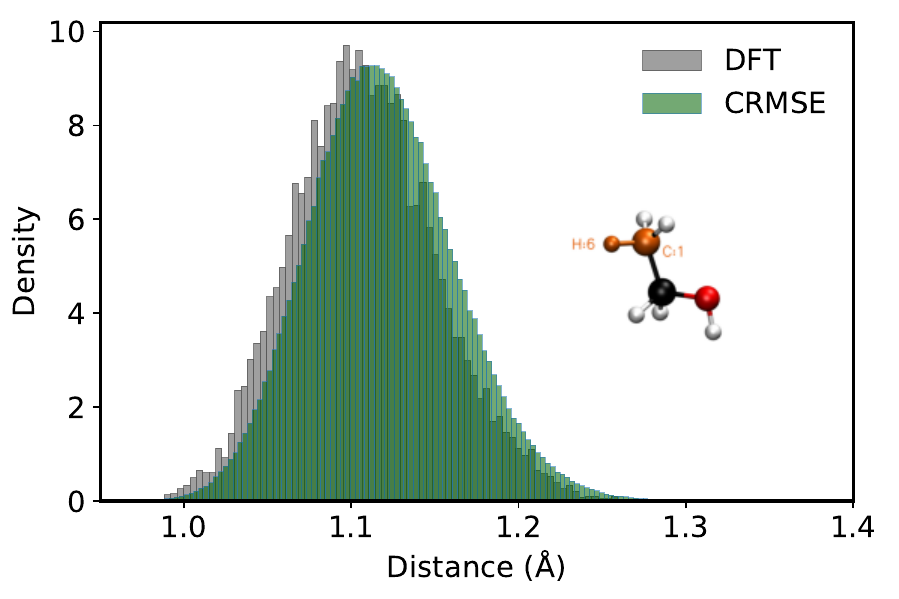}
        \caption{}
        \label{fig:crmse_dist_a}
    \end{subfigure}

    \vspace{0.3cm}

    \begin{subfigure}{\linewidth}
        \centering
        \includegraphics[width=\linewidth]{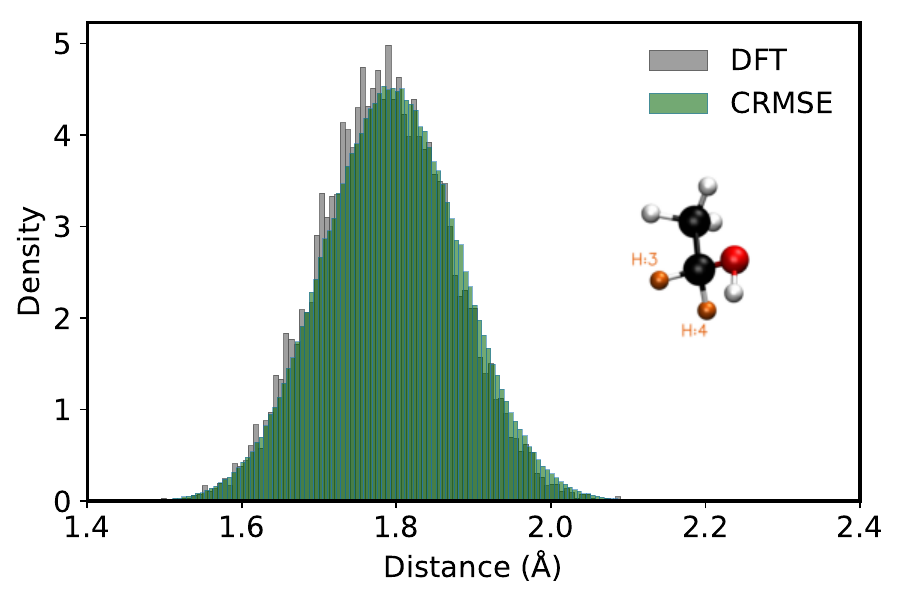}
        \caption{}
        \label{fig:crmse_dist_b}
    \end{subfigure}

    \vspace{0.3cm}

    \begin{subfigure}{\linewidth}
        \centering
        \includegraphics[width=\linewidth]{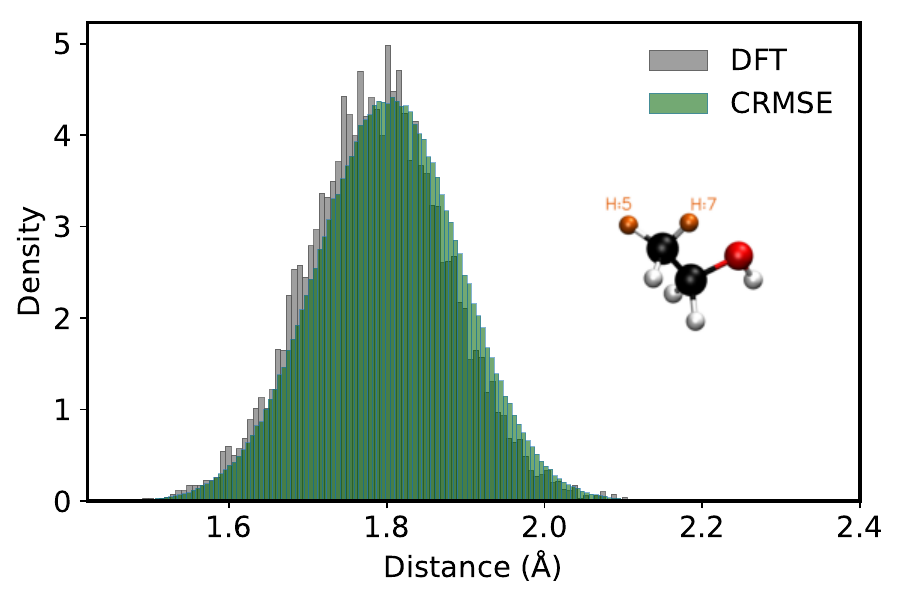}
        \caption{}
        \label{fig:crmse_dist_c}
    \end{subfigure}

    \caption{Densities of interatomic distances for selected atom pairs under CRMSE-MALA sampling (green) and the DFT reference (gray).}
    \label{fig:crmse_dist}
\end{figure}

We then incorporate the post-trained model into MALA under identical conditions to Sec.~\ref{sec:results_failure}. The chains now remain confined to the physical basin. Fig.~\ref{fig:crmse_dos} shows that the sampled energy distribution overlaps the DFT reference, and Fig.~\ref{fig:crmse_dist} confirms that the interatomic-distance distributions recover their physical form across all selected atom pairs, in contrast to the unphysical structures generated by the MSE model. The same correction holds for a second MD17 molecule, aspirin. The sampling conclusions are identical: the chains remain confined to the physical basin and the observables recover their DFT form, with the held-out forces likewise unchanged and the energies still within chemical accuracy, as detailed in Appendix~\ref{app:aspirin}.

\subsection{Mechanism: the replay-buffer energy distribution.}
\label{sec:results_mechanism}

\begin{figure}[ht!]
    \centering
    \includegraphics[width=1.\linewidth]{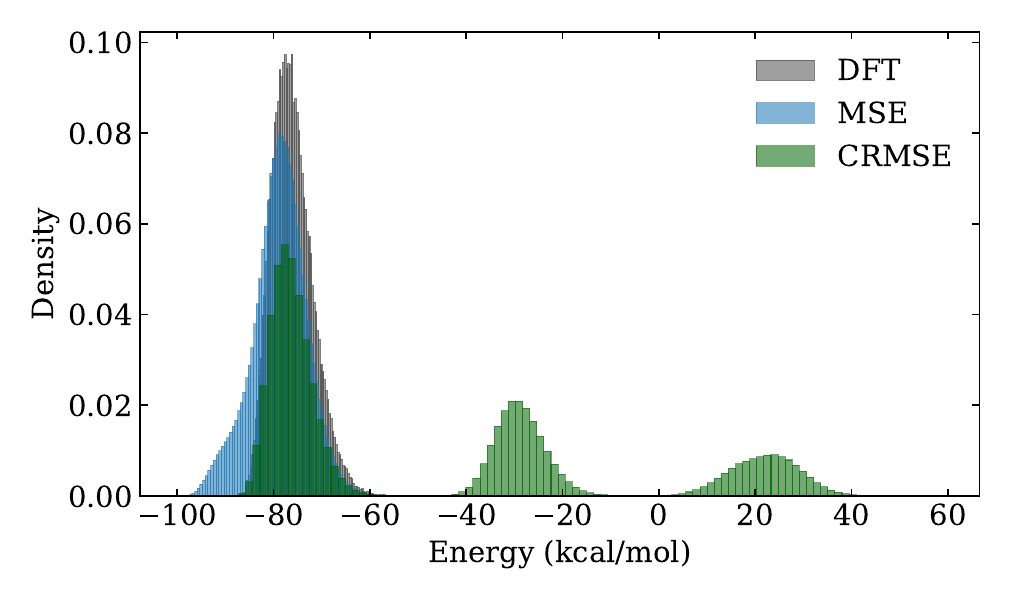}
    \caption{Energy distribution of the \emph{initial} replay-buffer configurations (those seeded into the buffer from the failed MSE-MALA chain, before any ULA refinement), evaluated under the MSE-trained model (blue) and under the CRMSE post-trained model (green). The density is computed on this fixed set of seed configurations, not on the evolved final buffer. Under CRMSE, the near-equilibrium configurations stay in the physical basin, while the unphysical ones are shifted to high energy, providing a direct measurement of the protective wall built by the contrastive term. The shaded region indicates the energy range of the DFT reference basin.}
    \label{fig:buffer_dos}
\end{figure}

To isolate the mechanism behind this correction, we examine how the two models score the replay-buffer configurations themselves, that is, the atypical states visited during the failed MSE-MALA chain. Fig.~\ref{fig:buffer_dos} shows the density of states of this fixed set of configurations evaluated under both potentials. Under the MSE potential, the buffer configurations extend below the physical basin, with a mean energy of $-78.8$ kcal/mol; this is precisely the spurious low-energy region that drew the sampler out of equilibrium. Under the CRMSE potential, the buffer configurations that are close to equilibrium remain in the physical basin, while the unphysical ones are raised far above it, into two groups near $-30$ and $+20$~kcal/mol; the mean energy of the buffer shifts from $-78.8$ to $-47.1$~kcal/mol ($\Delta = 31.7$~kcal/mol). This shift is a direct measurement of the protective energy wall posited in Sec.~\ref{sec:method}: the contrastive term raises the energy of the visited out-of-distribution configurations while leaving the physical basin intact, which is precisely what confines the corrected MALA trajectory and restores the observables of Sec.~\ref{sec:results_crmse}.

\subsection{Recovery of a thermodynamic observable}
\label{sec:results_freeenergy}

\begin{figure}
    \centering
    \begin{subfigure}{\linewidth}
        \centering
        \includegraphics[width=1\linewidth]{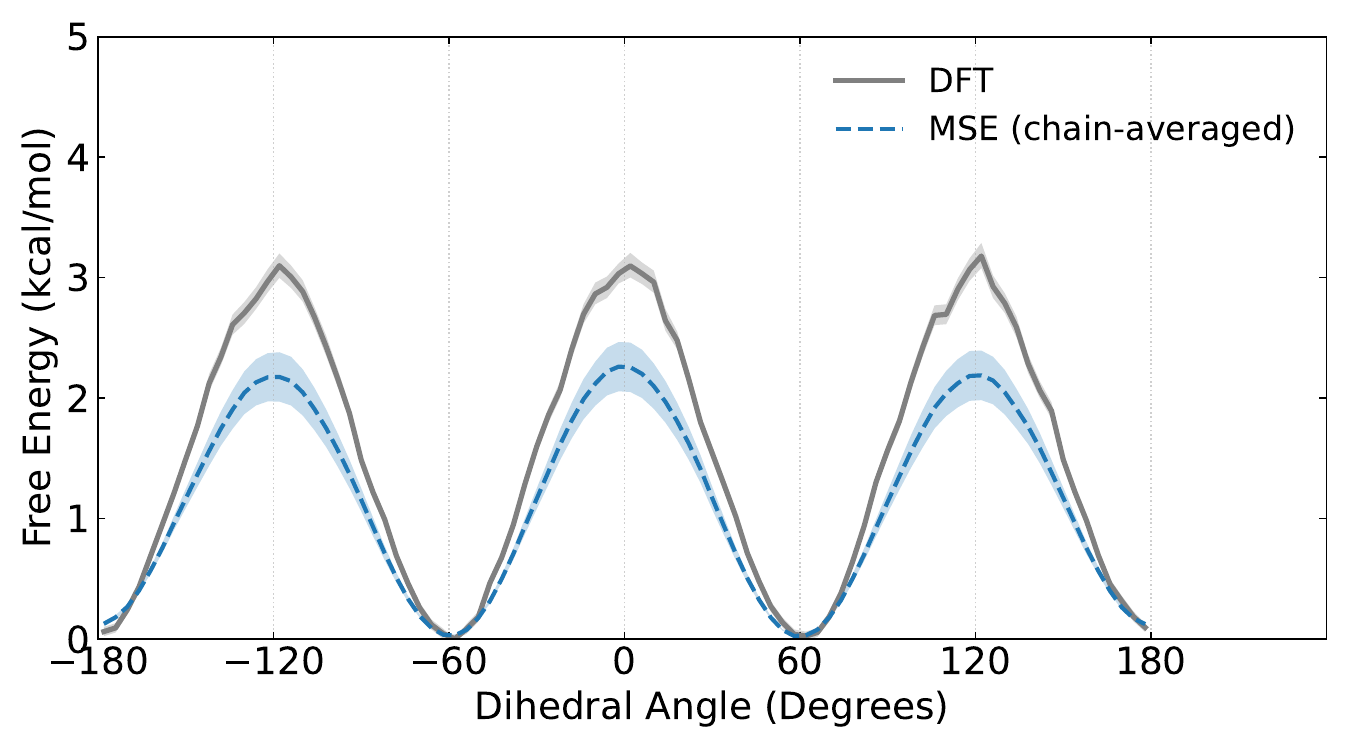}
        \caption{MSE-trained model}
        \label{fig:dihedral_mse}
    \end{subfigure}

    \vspace{0.3cm}

    \begin{subfigure}{\linewidth}
        \centering
        \includegraphics[width=1\linewidth]{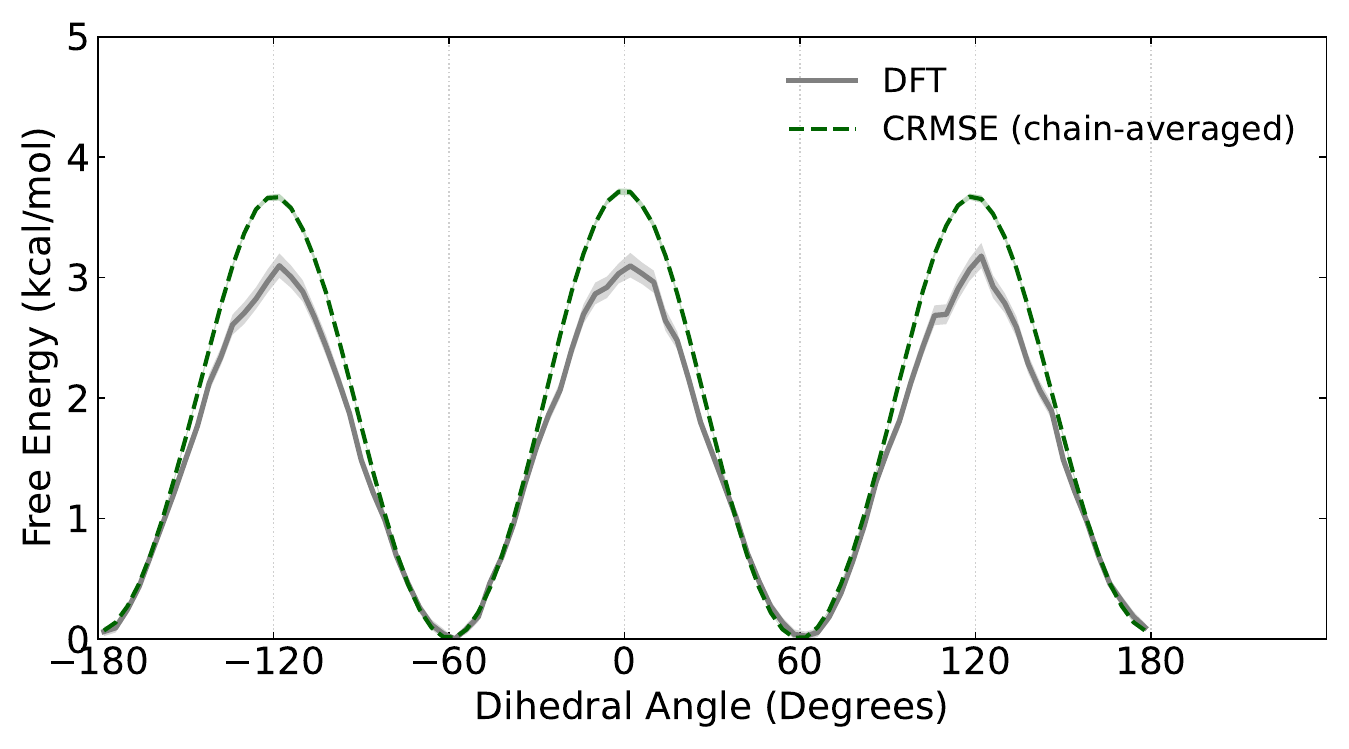}
        \caption{CRMSE post-trained model}
        \label{fig:dihedral_crmse}
    \end{subfigure}

    \caption{Free-energy profile along the $\mathrm{H\!-\!C\!-\!C\!-\!O\!}$ dihedral angle, chain-averaged over the $32$ independent MALA     chains ($\pm1$~SEM shaded). (a) The MSE model underestimates the torsional barriers and shows large inter-chain dispersion. (b) CRMSE recovers the barriers to near-quantitative agreement with the DFT reference, with dispersion reduced to within the line width.}
    \label{fig:dihedral}
\end{figure}

The agreement in the sampled energy distribution and interatomic distances establishes that CRMSE samples physically valid configurations, but does not by itself guarantee that the recovered ensemble has the correct relative populations. We therefore examine a thermodynamic observable that depends on the internal structure of the basin: the free-energy profile along the H$-$C$-$C$-$O dihedral angle $\phi$, $F(\phi) = -k_BT \ln p(\phi)$, computed at $T \approx 650$~K ($k_BT = 1.3$~kcal/mol) and averaged over the 32 independent MALA chains. Fig.~\ref{fig:dihedral} compares the MSE and CRMSE profiles against the DFT reference.

The MSE model systematically and substantially underestimates the torsional barriers, predicting peaks of $\approx$2.2~kcal/mol against the DFT value of $\approx$3.1~kcal/mol, and shows large inter-chain dispersion (shaded band: $\pm1$ standard error of the mean, SEM), the thermodynamic signature of the spurious low-energy regions that flatten the effective landscape. CRMSE recovers the barrier heights to near the reference value and, equally importantly, collapses the inter-chain dispersion to within the line width: the corrected chains agree both with the reference and with one another.

A small residual remains: near the barrier maxima ($\phi \approx -120^\circ, 0^\circ, 120^\circ$) CRMSE slightly overestimates the free energy by $\approx$0.5$-$0.6~kcal/mol. These high-energy angles are sparsely sampled in both the training data and the DFT reference, whose own statistical uncertainty there is shown as the band in Fig.~\ref{fig:dihedral}. We do not claim this residual is an improvement; we note only that it is small and opposite in sign to the substantially larger MSE error.

\subsubsection{Robustness to reduced training data}
\label{sec:thermo_200}

To assess whether CRMSE remains effective when the labeled dataset is
substantially reduced, we repeat the ethanol experiment starting from a
training set of only 200 DFT configurations. This is well below the
maximum of 1000 structures recommended for rMD17 (Sec.~\ref{sec:results_failure}), so this experiment probes the regime
of genuine data scarcity rather than mere subsampling. Full details of the
training procedure and the corresponding energy and force accuracies are
reported in Appendix~\ref{app:appendixE}.

\begin{figure}
    \centering
    \begin{subfigure}{\linewidth}
        \centering
        \includegraphics[width=1\linewidth]{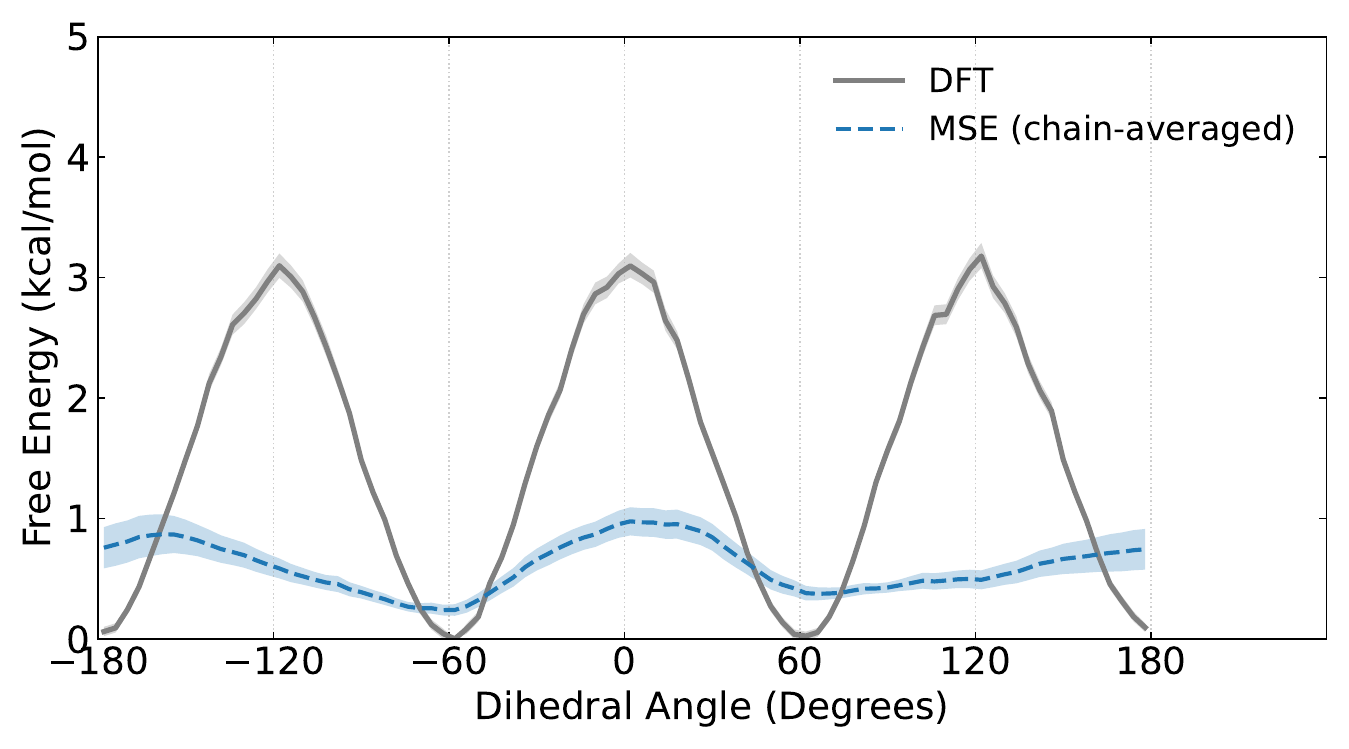}
        \caption{MSE-trained model}
        \label{fig:dihedral_200_mse}
    \end{subfigure}
    \vspace{0.3cm}
    \begin{subfigure}{\linewidth}
        \centering
        \includegraphics[width=1\linewidth]{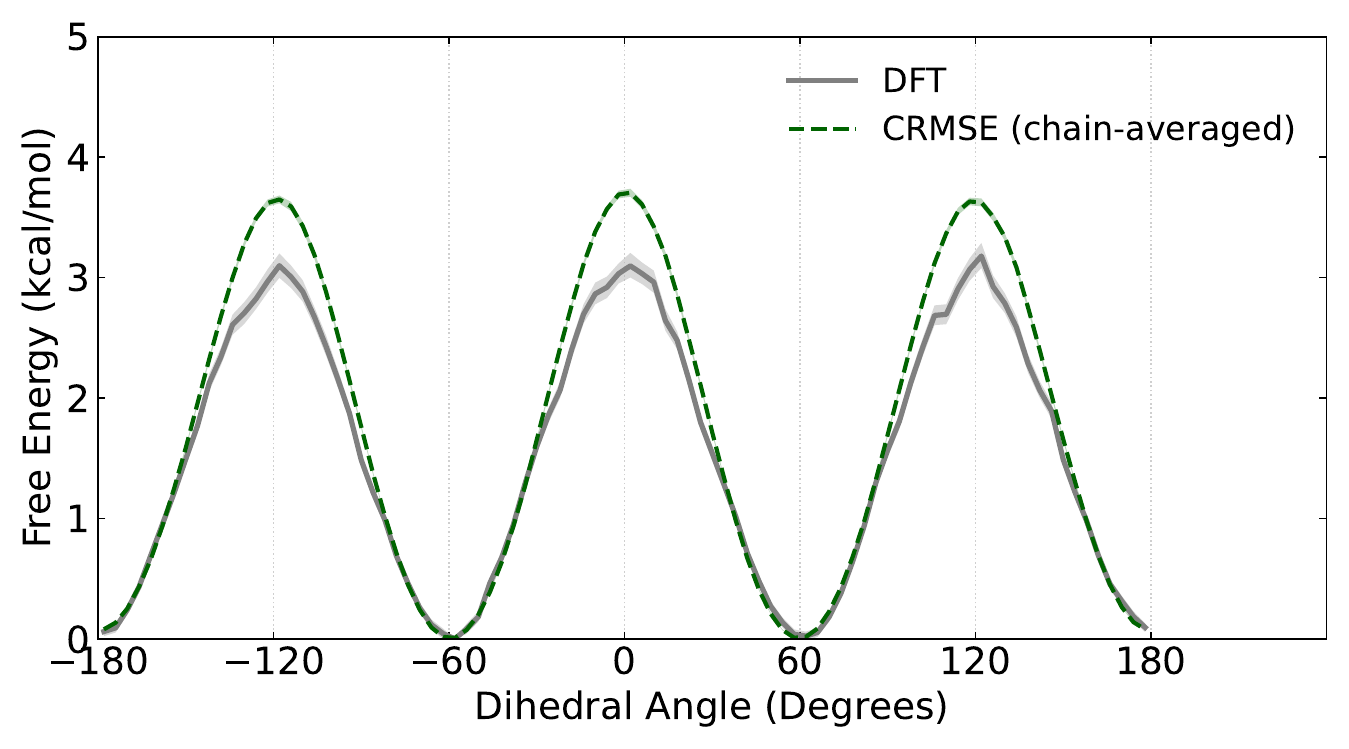}
        \caption{CRMSE post-trained model}
        \label{fig:dihedral_200_crmse}
    \end{subfigure}
    \caption{Free-energy profile along the $\mathrm{H\!-\!C\!-\!C\!-\!O}$ dihedral angle for the 200-configuration training set, chain-averaged over the $32$ independent MALA chains ($\pm1$~SEM shaded). (a) With little training data the MSE profile is badly distorted: it is strongly asymmetric and the torsional barriers are much too low. (b) CRMSE recovers the barrier heights and the shape of the profile, almost matching the full-data CRMSE result of Fig.~\ref{fig:dihedral}b.}
    \label{fig:dihedral_200}
\end{figure}

We perform MALA simulations with both the data-limited MSE and CRMSE
potentials and compare the resulting free-energy profiles against the
DFT reference, as shown in Fig.~\ref{fig:dihedral_200}. The MSE model
produces a severely distorted profile, strongly asymmetric and with torsional barriers substantially underestimated, which reflects the compounded effect of data scarcity on the learned energy landscape. The
CRMSE model, by contrast, recovers barrier heights and profile shape in
close agreement with the DFT reference, and consistent with the CRMSE
result obtained using the full training dataset. These results highlight
the key advantage of CRMSE over standard MSE training: even when the
labeled dataset is significantly reduced, the contrastive replay
mechanism enables the model to recover configurations whose thermodynamic
properties match the DFT reference, demonstrating improved data
efficiency and robustness to training set size.

\subsubsection{Generalization to aspirin}
\label{sec:thermo_aspirin}

\begin{figure}
    \centering
    \begin{subfigure}{\linewidth}
        \centering
        \includegraphics[width=1\linewidth]{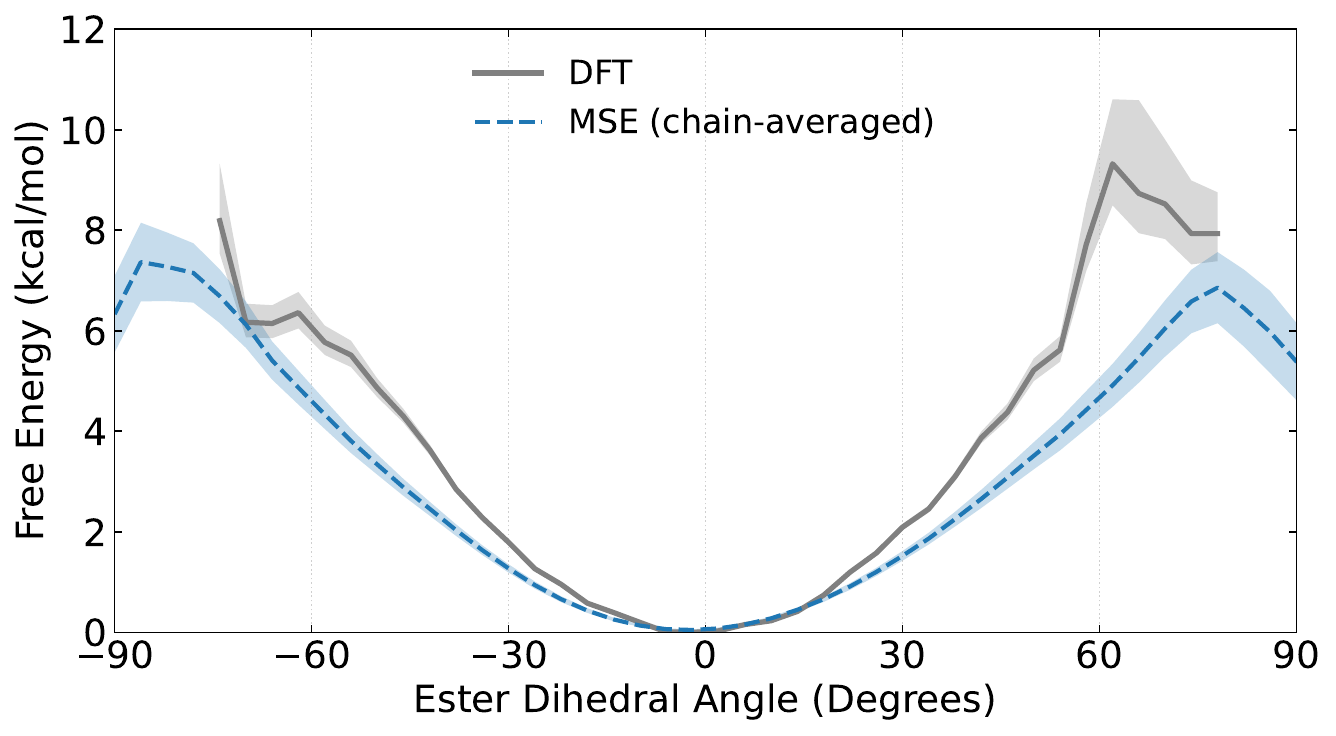}
        \caption{MSE-trained model}
        \label{fig:ester_mse}
    \end{subfigure}
    \vspace{0.3cm}
    \begin{subfigure}{\linewidth}
        \centering
        \includegraphics[width=1\linewidth]{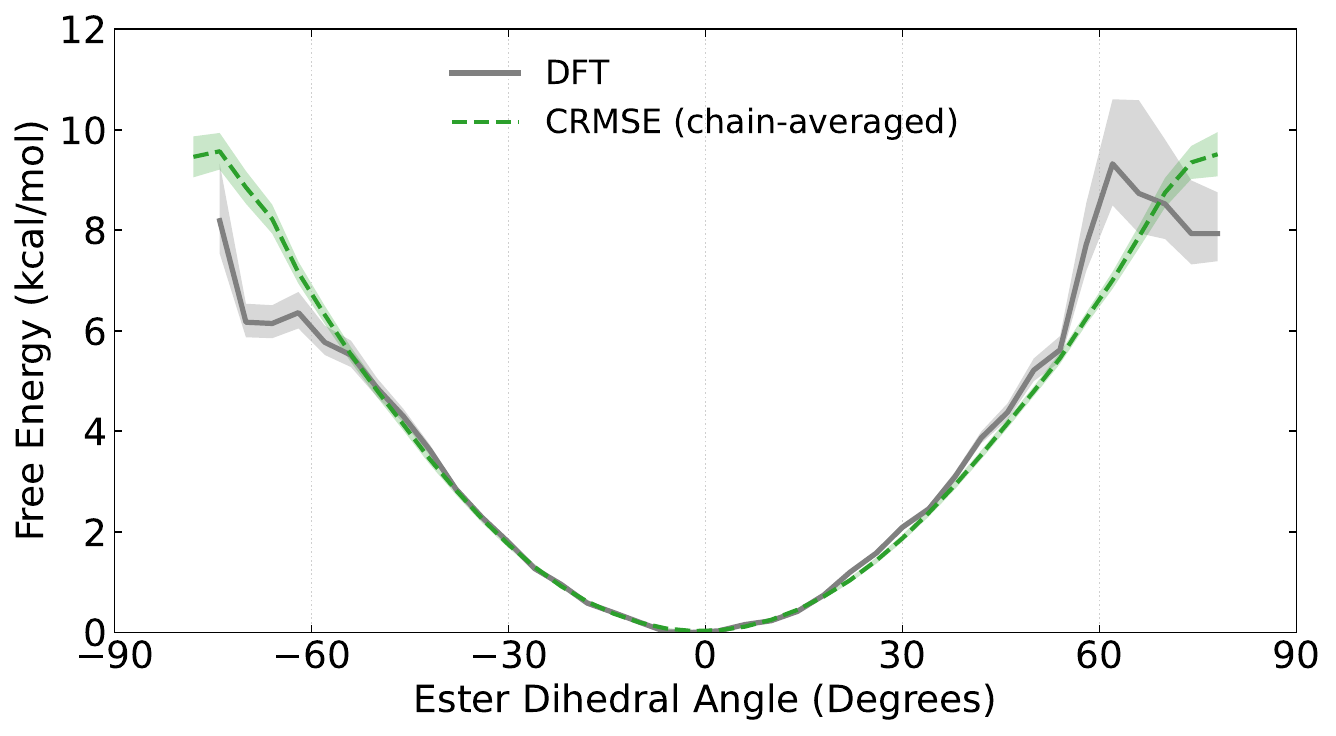}
        \caption{CRMSE post-trained model}
        \label{fig:ester_crmse}
    \end{subfigure}
    \caption{Free-energy profile along the ester dihedral $\psi$ of aspirin, chain-averaged over the $32$ independent MALA chains ($\pm1$~SEM shaded). (a) The MSE model reproduces the bottom of the well but rises too slowly away from it, lying $1$--$2$~kcal/mol below the DFT reference for $40^\circ \lesssim |\psi| \lesssim 65^\circ$ (up to ${\sim}4$~kcal/mol near $\psi \approx +60^\circ$). (b) CRMSE follows the DFT reference within its statistical uncertainty for $|\psi| \lesssim 55^\circ$; beyond, where the DFT data are sparse, the two differ by up to ${\sim}2.5$~kcal/mol in either direction.}
    \label{fig:ester}
\end{figure}

Having established the validity of CRMSE on ethanol, we now examine whether the same correction generalises to a more complex molecule. As for ethanol, the surrogate is trained on 950 aspirin configurations, the maximum recommended for rMD17 (Sec.~\ref{sec:results_failure}; full training details in Appendix~\ref{app:aspirin}). Aspirin introduces a torsion of a different kind: the ester group ($\mathrm{C\!-\!O\!-\!C\!=\!O}$ dihedral, hereafter $\psi$; C$_{6}$--O$_{12}$--C$_{11}$=O$_{8}$ in Fig.~\ref{fig:dft_aspirin}). The partial double-bond character of the ester C--O bond keeps the group in its planar $Z$ (cis) conformation, $\psi \approx 0^\circ$, and the reference trajectory never leaves this well, so its free-energy profile tests how steeply the free energy rises as the ester twists out of plane. We evaluate the free-energy profile $F(\psi) = -k_BT \ln p(\psi)$ at $T \approx 580$~K ($k_BT = 1.155$~kcal/mol), averaged over the $32$ independent MALA chains. Fig.~\ref{fig:ester} compares the MSE and CRMSE profiles against the DFT reference.

The MSE model reproduces the bottom of the well of the ester-torsion profile ($|\psi| \lesssim 30^\circ$), but beyond it the free energy rises too slowly: for $40^\circ \lesssim |\psi| \lesssim 65^\circ$ it lies $1$--$2$~kcal/mol below the DFT reference, and up to about $4$~kcal/mol below near $\psi \approx +60^\circ$, so twisted ester conformations become too easy to reach. The spread between chains (shaded $\pm1$~SEM band) also grows in this region.

CRMSE corrects this. For $|\psi| \lesssim 55^\circ$ the chain-averaged CRMSE profile lies within the DFT statistical uncertainty, and the spread between chains shrinks to about the line width. Beyond that, where the DFT reference rests on few configurations, the two curves differ by up to about $2.5$~kcal/mol in either direction: CRMSE lies above the reference near $\psi \approx -70^\circ$ and below it near $\psi \approx +60^\circ$.

\subsubsection{Direct applicability to molecular dynamics: BAOAB Langevin dynamics sampling}
\label{sec:ula_validation}

We also run Langevin dynamics with the BAOAB integrator~\cite{leimkuhler2013rational}, which follows the inertial motion of the atoms (no overdamped limit is taken). Our aim is to test whether CRMSE also works in physical simulations that, unlike MCMC methods such as the MALA used above, have no accept/reject step to keep them stable~\cite{Roberts1996Langevin}. For both potentials, we run 10 independent chains of $10^7$ steps each ($\Delta t = 0.5$~fs, $\gamma = 1\,\mathrm{ps}^{-1}$, $k_BT = 1.3$~kcal/mol), started from DFT configurations with Maxwell--Boltzmann velocities and using the physical atomic masses, and compute the free-energy profiles of the $\mathrm{H\!-\!C\!-\!C\!-\!O}$ methyl rotor and of the O--H (hydroxyl) rotor.

\begin{figure}
    \centering
    \begin{subfigure}{\linewidth}
        \centering
        \includegraphics[width=1\linewidth]{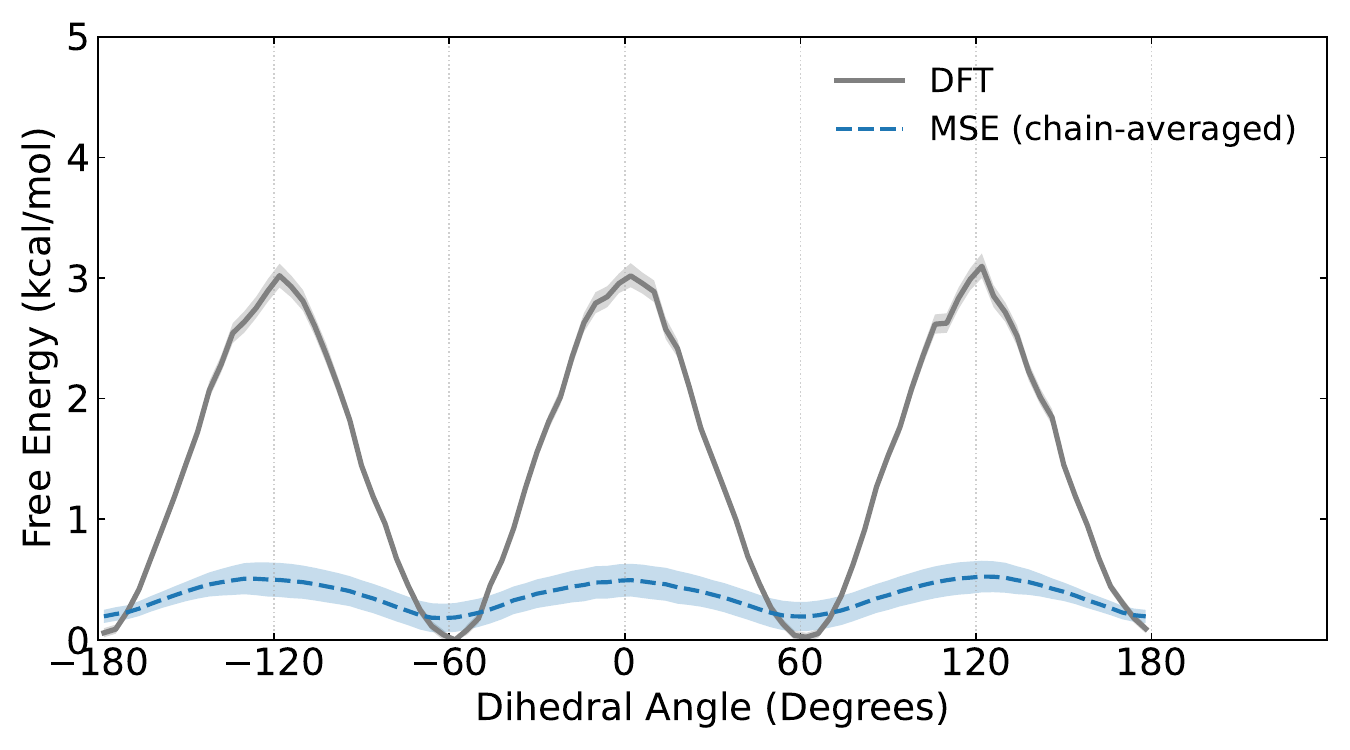}
        \caption{MSE-trained model}
        \label{fig:baoab_methyl_mse}
    \end{subfigure}

    \vspace{0.3cm}

    \begin{subfigure}{\linewidth}
        \centering
        \includegraphics[width=1\linewidth]{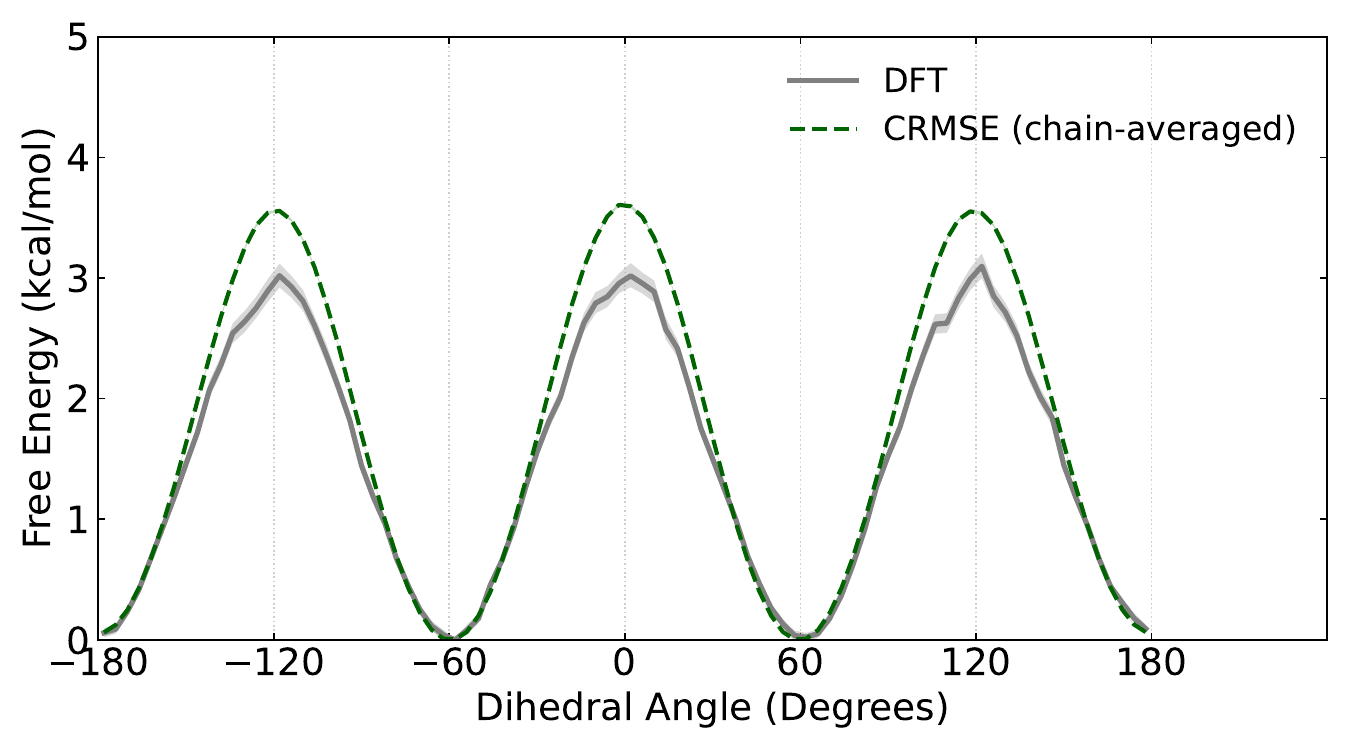}
        \caption{CRMSE post-trained model}
        \label{fig:baoab_methyl_crmse}
    \end{subfigure}

    \caption{Free-energy profile along the $\mathrm{H\!-\!C\!-\!C\!-\!O}$ methyl-rotor dihedral under BAOAB Langevin dynamics, chain-averaged over $10$ independent chains of $5$~ns each ($\pm1$~SEM shaded; for CRMSE the band is narrower than the line width). (a) With the MSE model the torsional barrier almost disappears. (b) CRMSE recovers the barrier, slightly overestimated.}
    \label{fig:baoab_methyl}
\end{figure}

Fig.~\ref{fig:baoab_methyl} shows the methyl-rotor profile. With the MSE potential the torsional barrier almost disappears. With the CRMSE potential the barrier is recovered, slightly overestimated.

The hydroxyl rotor shows the same picture (Fig.~\ref{fig:baoab_oh}, Appendix~\ref{app:oh_rotor}): under BAOAB the MSE profile moves far from the DFT reference, while CRMSE stays close to it. Even after $10^7$ steps, ten times longer than the MALA chains, CRMSE keeps sampling correctly and all its chains remain in the physical basin, while the MSE potential does not.

\subsection{Comparison with a static geometric prior}
\label{sec:results_prior_reweighting}   %

CRMSE restores correct sampling by changing the learned energy landscape. A common alternative leaves the potential as it is and restrains the sampler from outside, with a fixed geometric prior that keeps the internal coordinates within the range seen in the training data, so that the dynamics cannot drift into unphysical geometries. We test whether such a prior, added to the uncorrected MSE potential of Sec.~\ref{sec:results_failure} (not to CRMSE), is enough on its own to recover the correct thermodynamics. As is common in bottom-up coarse-graining, the prior consists of soft harmonic walls on the bond lengths $d$ and valence angles $\alpha$, placed at the smallest and largest values found in the training configurations,

\begin{equation}\begin{split}
U_{\text{prior}}(\mathbf{S}) ={}& k_{d} \sum_{b} \max\!\big(0,\, d_b - (d_b^{\max}+\delta_d)\big)^2 \\
&+ k_{d} \sum_{b} \max\!\big(0,\, (d_b^{\min}-\delta_d) - d_b\big)^2 \\
&+ k_{\alpha} \sum_{a} \max\!\big(0,\, \alpha_a - (\alpha_a^{\max}+\delta_\alpha)\big)^2 \\
&+ k_{\alpha} \sum_{a} \max\!\big(0,\, (\alpha_a^{\min}-\delta_\alpha) - \alpha_a\big)^2
\end{split}
\label{eq:prior}
\end{equation}

where $d_b^{\min/\max}$ and $\alpha_a^{\min/\max}$ are the smallest and largest bond lengths and angles in the training data, $\delta_d,\delta_\alpha$ are small tolerances, and $k_d,k_\alpha$ set the stiffness of the walls. We then sample the combined potential $U_\theta(\mathbf{S}) = U_\theta^{\text{model}}(\mathbf{S}) + U_{\text{prior}}(\mathbf{S})$.

We ran 32 independent MALA chains with this prior, using the stiffness $(k_d,k_\alpha)=(5000,5000)$ chosen by the scan in Appendix~\ref{app:prior_decomposition}, and computed the chain-averaged $\mathrm{H\!-\!C\!-\!C\!-\!O}$ dihedral profile as we did for CRMSE. Even with the prior, some chains still leave the physical geometry. Appendix~\ref{app:prior_decomposition} shows that the chains that stay within the prior's bounds give a barrier close to the CRMSE one, while those that cross the bounds give a barrier close to that of the uncorrected MSE potential.

\begin{figure}[h]
    \centering
    \includegraphics[width=1\linewidth]{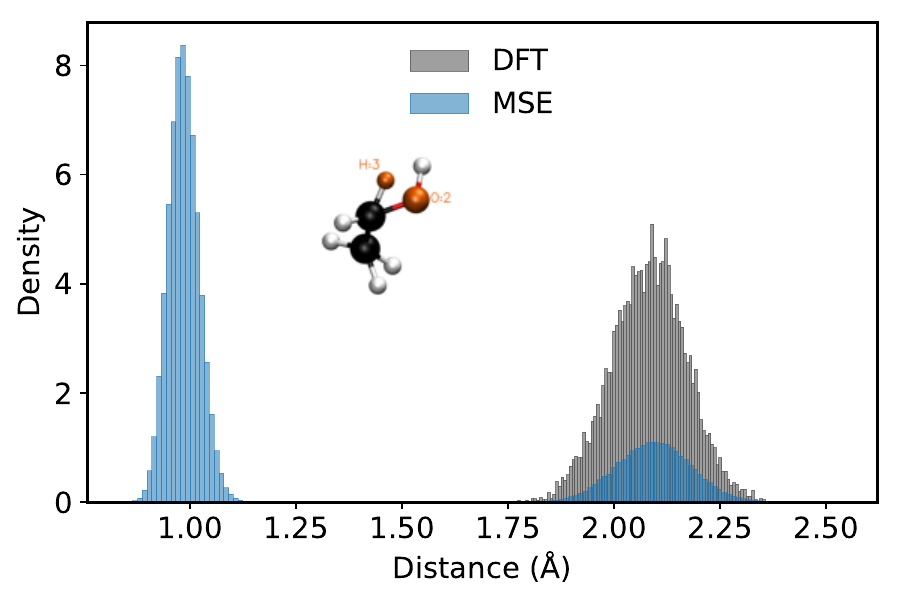}
    \caption{Distribution of a selected O--H distance for one of the off-target chains of Appendix~\ref{app:prior_decomposition}, sampled with the MSE potential plus the geometric prior. A separate group of frames lies beyond the physical range.}
    \label{fig:oh_prior_chain}
\end{figure}

Fig.~\ref{fig:oh_prior_chain} shows this directly, with the O--H distance distribution of one of the chains that leaves the physical geometry.

A static geometric prior is therefore not enough on its own. Whether a chain gives the CRMSE-like or the MSE-like barrier depends on whether it happens to stay inside the bounds, and the prior cannot guarantee that. CRMSE keeps all $32$ chains in the physical basin (Sec.~\ref{sec:results_crmse}) because it changes the landscape itself instead of fencing it off.

\section{Conclusion}
\label{sec:conclusion}

We have shown that machine learning interatomic potentials
trained by standard MSE minimization can be accurate by
every regression metric yet fail catastrophically as
samplers: with energies and forces at chemical accuracy on
held-out DFT data, their trajectories, under both MALA and uncorrected Langevin dynamics, nonetheless escape the physical basin into spurious low-energy minima
and return thermodynamic observables that bear no
resemblance to the reference. This failure is not an
artifact of a particular architecture but a statistical
inevitability of regression on finite data in high
dimensions, where the training set constrains only a
vanishing fraction of configuration space.

To correct it, we introduced CRMSE, a cheap post-training
step that augments the MSE with a contrastive term derived
from the Kullback--Leibler divergence between the
potential's implicit Boltzmann distribution and the target.
The network serves as its own energy-based model:
persistent Langevin chains expose the configurations it
drifts into and raise their energy, building a protective
barrier around the physical basin without any new ab initio
data or change of architecture. On ethanol and aspirin from
MD17, CRMSE preserves force accuracy and keeps energy errors within chemical accuracy while recovering the sampled energy distribution, the interatomic-distance distributions, and the
dihedral free-energy profiles to near-quantitative
agreement with DFT, whereas the MSE model misses these observables by large margins. The correction also carries over to molecular dynamics with inertia. With the BAOAB integrator, the methyl-rotor profile is recovered with the same small residual as with MALA (Sec.~\ref{sec:ula_validation}), and the hydroxyl rotor, a second torsion not used to tune the method, agrees with DFT better than any other dihedral in this work (Appendix~\ref{app:oh_rotor}). The inter-chain dispersion, a thermodynamic
signature of the spurious landscape, collapses to within
the line width, and the correction remains effective when
the training set is reduced from 950 to 200 configurations
(Appendix~\ref{app:appendixE}) and across almost two orders of magnitude in the regularization strength
$\lambda$ (Appendix~\ref{app:lambda}).

CRMSE is complementary to active learning rather than a
replacement for it: it restores reliable sampling by
exploiting the Boltzmann normalization, but does not
improve the potential's accuracy in the off-manifold
regions, which still demands fresh data.
Hand-designed geometric priors are another option: they fence the sampler in instead of reshaping the landscape. We find (Sec.~\ref{sec:results_prior_reweighting}) that they do not reliably restore correct sampling, even when carefully tuned. Some chains stay inside the bounds and give a CRMSE-like profile, while others escape and fall back to the uncorrected result. CRMSE avoids this because it removes the spurious minima from the potential itself.
What we find most
significant is how little the correction requires. Applied
to one of the most widely used benchmarks in the field,
with a standard architecture and no new ab initio data, it
repairs a potential already well-optimized by every
regression metric. This is evidence that reliable sampling depends
less on the volume of data than on training the potential
to reproduce the distribution it is sampled from, something
the regression objective structurally omits. Because every
evaluation is performed under the learned potential rather
than the ab initio reference, the correction should carry over to large systems at modest cost, precisely where extending
the reference data is hardest and sampling failures are
most consequential.

The lesson is not special to chemistry: any surrogate later
used as a generator should be trained the way it is used: to reproduce its own sampling distribution, not
merely to fit a finite set of points. The natural next
steps are larger and more complex systems, such as proteins
and periodic crystalline materials, where configuration
space grows, spurious minima proliferate, and reliable
Boltzmann sampling is most critically needed.

\FloatBarrier

\begin{acknowledgments}
We thank Iakovos Evdaimon for useful comments.
The authors used Claude (Claude Opus 5.5, Anthropic) to improve the clarity, grammar and wording of the text, and to refine analysis scripts written by the authors. All suggestions were reviewed and verified by the authors, who take full responsibility for the content of this article.
\end{acknowledgments}

\section*{Data and code availability}
The (revised) MD17 dataset analysed in this work is publicly available
(Refs.~\cite{Christensen2020rMD17,Christensen2020}). The code implementing the CRMSE
post-training procedure and the scripts required to reproduce the
figures of this paper are openly available on Zenodo, doi:10.5281/zenodo.23371888~\cite{Tzivrailis2026Zenodo}.

\appendix
\appendix

\section{Sensitivity to CRMSE hyperparameters}
\label{app:lambda}

CRMSE introduces one main hyperparameter beyond those of
the underlying MSE training: the regularization strength
$\lambda$ of Eq.~\eqref{eq:crmse}. We assess its
effect by varying $\lambda$ while keeping the remaining
parameters fixed at the values used in the main text
($K = 200$ ULA steps per update, step size $\eta =
0.0004$), and report the Kolmogorov--Smirnov (KS) statistic between the sampled and reference energy distributions for each setting, which quantifies sampling fidelity (lower is better).
\begin{figure}[H]
        \centering
        \includegraphics[width=0.7\linewidth]{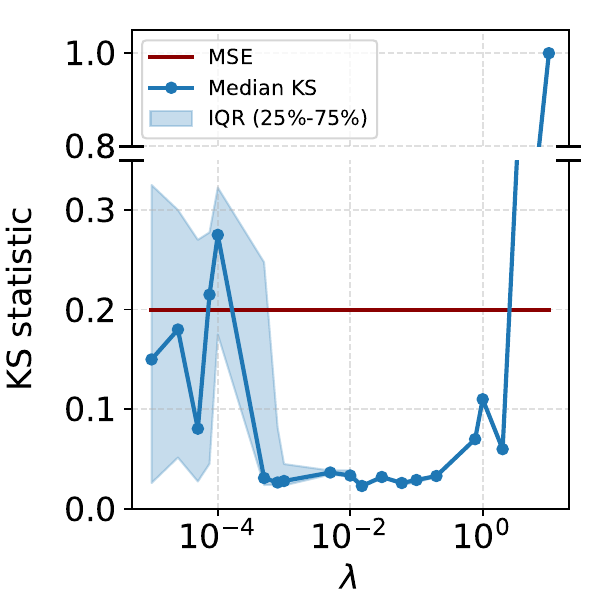}
    \caption{Sensitivity of CRMSE to the regularization
            strength $\lambda$, assessed under the protocol of
            Appendix~\ref{app:protocol} with $K = 200$ ULA steps
            and step size $\eta = 0.0004$ fixed. The $y$-axis
            reports the Kolmogorov--Smirnov (KS) statistic between
            the sampled and reference energy distributions, with
            lower values indicating better agreement. For small
            $\lambda$, the contrastive term is too weak to suppress
            the spurious minima, resulting in large KS values and
            high inter-chain variance. Over an intermediate range
            the method achieves its best sampling fidelity, with
            the KS statistic reaching a stable minimum. For large
            $\lambda$, the contrastive push begins to distort the
            data-constrained basin, and the KS statistic rises
            again. The dashed vertical line indicates the operating
            value used in the main text.}
    \label{fig:hyperparams}
\end{figure}
\subsection{Evaluation protocol}
\label{app:protocol}

For the $\lambda$ hyperparameter setting we perform the full CRMSE
post-training starting from the same MSE-pretrained weights and optimizer state.
We vary $\lambda$ about the operating point used in the main text
($\lambda = 0.015$), while keeping the ULA step size $\eta = 0.0004$ and the
number of steps per update $K = 200$ fixed at their main-text values. The
results are summarized in Fig.~\ref{fig:hyperparams}.

\subsection{Regularization strength $\lambda$}
\label{app:sweep_lambda}

Fig.~\ref{fig:hyperparams} shows the effect of $\lambda$, varied over
[$10^{-5}$, 10] on a logarithmic scale. The KS statistic shows a broad stable plateau. For small $\lambda$ the contrastive term is too
weak to raise the energy of the out-of-distribution configurations, and the KS statistic
remains large as the sampler continues to escape the physical basin. For large $\lambda$ the contrastive push begins to distort the data-constrained basin and the KS statistic rises again. Between these regimes lies a stable window spanning almost two orders of magnitude, $\lambda \in [0.01, 0.5]$, over which the KS statistic is small. The operating value $\lambda = 0.015$ lies within this window. The width of the plateau demonstrates that the correction is robust and does not rely on fine-tuning of $\lambda$.

\section{Preconditioned MALA sampling}
\label{app:precond}

All Markov chain Monte Carlo samples in this study are obtained with a
preconditioned variant of the Metropolis-Adjusted Langevin Algorithm (MALA).
Given the current configuration $\mathbf{x} \in \mathbb{R}^{3n}$ and a diagonal preconditioning matrix $\mathbf{H} \in \mathbb{R}^{3n \times 3n}$, a
candidate configuration $\mathbf{x'}$ is proposed by the following equation :
\begin{equation}
    \mathbf{x}' = \mathbf{x} - \mathbf{H}\, \nabla_{\mathbf{x}}
    E(\mathbf{x}) + \sqrt{2\mathbf{H}k_BT}\, \boldsymbol{\xi}
    \label{eq:mala_preconditioned}
\end{equation}
where $\sqrt{2\mathbf{H}k_BT}$ denotes the elementwise square root of the
diagonal matrix $2\mathbf{H}k_BT$, and $\boldsymbol{\xi} \sim
\mathcal{N}(\mathbf{0},\,\mathbf{I}_{3n})$. So, the proposal distribution :

\begin{equation}
    q_H(\mathbf{x}'|\mathbf{x}) = \mathcal{N}\!\left(\mathbf{x} - \mathbf{H}\,
    \nabla_{\mathbf{x}} E(\mathbf{x}),\; 2k_B T\mathbf{H}\,\right)
\end{equation}

\begin{align}
    q_H(\mathbf{x}'|\mathbf{x})
    &= \frac{1}{(2\pi)^{3n/2} \left|2k_BT\,\mathbf{H}\right|^{1/2}} \notag\\
    &\quad\times\exp\!\left( -\frac{1}{2}(\mathbf{x}' - \mathbf{x}
    + \mathbf{H}\,\nabla_{\mathbf{x}} E(\mathbf{x}))^\top \right.\notag\\
    &\qquad\quad\left. \times\left(2k_BT\,\mathbf{H}\right)^{-1}
    (\mathbf{x}' - \mathbf{x} + \mathbf{H}\,\nabla_{\mathbf{x}} E(\mathbf{x})) \right)
\end{align}

Since $\mathbf{H} = \varepsilon\times\operatorname{diag}(H_{11},\dots,H_{3n,3n})$, with $\varepsilon$ the global step size, the determinant
and the inverse are diagonal too, and so the proposal distribution takes the explicit form:

\begin{multline}
    q_H(\mathbf{x}'|\mathbf{x}) = \frac{1}{(4\pi k_BT\varepsilon)^{3n/2}
    \prod_{i=1}^{3n}\sqrt{H_{ii}}} \\
    \times\exp\!\left(-\frac{1}{4k_BT\varepsilon}\sum_{i=1}^{3n}
    \frac{\left(x'_i - x_i - \varepsilon H_{ii}F_i(\mathbf{x})\right)^2}{H_{ii}}\right)
    \label{eq:mala_preconditioned_proposal explicit}
\end{multline}

Finally the Metropolis--Hastings acceptance probability :

    \begin{equation}
        \alpha(\mathbf{x}, \mathbf{x}') = \min\!\left(1,\,
        \frac{p(\mathbf{x}')\, q_H(\mathbf{x}|\mathbf{x}')}{p(\mathbf{x})\,q_H(\mathbf{x}'|\mathbf{x})}
        \right)
    \end{equation}

where $p(\mathbf{x}) \propto e^{-\beta E(\mathbf{x})}$ is the target Boltzmann
distribution. And so the acceptance probability :

\begin{align}
\alpha(\mathbf{x},\mathbf{x}')
    &= \min\Biggl(1,\; \exp\Biggl[ \notag\\
    &\quad -\frac{E(\mathbf{x}')-E(\mathbf{x})}{k_B T} \notag\\
    &\quad +\frac{1}{4 k_B T\varepsilon} \sum_{i=1}^{3n} \frac{1}{H_{ii}} \notag\times\Bigl[ \bigl(x'_i - x_i - \varepsilon H_{ii}F_i(\mathbf{x})\bigr)^2 \notag\\
    &\qquad\phantom{\times\Bigl[}
        -\bigl(x_i - x'_i - \varepsilon H_{ii}F_i(\mathbf{x}')\bigr)^2
    \Bigr]\Biggr]\Biggr)
    \label{eq:mala_preconditioned_acceptance explicit}
\end{align}

The core idea behind this modification is to incorporate in the algorithm the essence of the mass of each different atom as a measure of its inertia. Since heavier atoms move less than lighter ones, weights of the precondition matrix are set inversely proportional to the respective mass of the atom they are acted on. Also, higher preconditioned weights are attributed to atoms of the same type that can move with more freedom due to the specific structure of the molecule (e.g., the hydroxyl hydrogen in ethanol). In that way we can obtain a more effective exploration of the multidimensional space of the target probability distribution $\Pi(\mathbf{x})$ due to the fact that we mimic better the true behavior of the physical system.

\begin{table}[h!]
    \centering
    \caption{Per-atom preconditioning weights $H_{aa}$ used in
    Eq.~\eqref{eq:mala_preconditioned_proposal explicit} and in
    Eq.~\eqref{eq:mala_preconditioned_acceptance explicit} for the MALA
    simulations in the main text.}
    \label{tab:precond_weights}
    \begin{ruledtabular}
    \begin{tabular}{llc}
        Species & Chemical environment & Weight $H_{aa}$ \\
        \colrule
        C  & CH$_3$ carbon          & 0.34 \\
        C  & CH$_2$ carbon          & 0.36 \\
        O  & Hydroxyl oxygen        & 0.32 \\
        H  & CH$_3$ hydrogens       & 1.45 \\
        H  & CH$_2$ hydrogens       & 1.50 \\
        H  & Hydroxyl hydrogen (OH) & 1.85 \\
    \end{tabular}
    \end{ruledtabular}
\end{table}

Since the configuration space of the molecule is isotropic, we assign a
single preconditioning weight $H_{aa}$ per atom rather than separate weights
for each Cartesian direction. The resulting values for each atom in the
ethanol molecule are listed in Table~\ref{tab:precond_weights}. With this
choice, the step size is fixed at $\varepsilon = 0.0005$ for all production
runs, yielding an average acceptance rate of approximately $55\%$, consistent
with the regime of efficient MALA sampling.

\begin{table}[h!]
    \centering
    \caption{Per-atom preconditioning weights $H_{aa}$ used in
    Eq.~\eqref{eq:mala_preconditioned_proposal explicit} and in
    Eq.~\eqref{eq:mala_preconditioned_acceptance explicit} for the MALA
    simulations of aspirin in the main text.}
    \label{tab:precond_weights_aspirin}
    \begin{ruledtabular}
    \begin{tabular}{llc}
        Species & Chemical environment & Weight $H_{aa}$ \\
        \colrule
        C & Aromatic ring carbon              & 0.38 \\
        C & Carboxyl carbon                   & 0.43 \\
        C & Acetyl carbonyl carbon            & 0.44 \\
        O & Carboxyl carbonyl oxygen          & 0.34 \\
        O & Carboxyl hydroxyl oxygen          & 0.36 \\
        O & Ester oxygen                      & 0.36 \\
        O & Acetyl carbonyl oxygen            & 0.34 \\
        H & Carboxylic acid hydrogen          & 1.80 \\
        H & Aromatic hydrogen                 & 1.40 \\
        H & Acetyl methyl hydrogen            & 1.55 \\
        H & Additional hydrogen              & 1.45 \\
    \end{tabular}
    \end{ruledtabular}
    \vspace{0.2cm}

\end{table}

For the aspirin molecule, the same preconditioning
scheme is applied, with per-atom weights listed in
Table~\ref{tab:precond_weights_aspirin}. The global
step size is set to $\varepsilon = 0.0006$, yielding
an average acceptance rate of approximately $55\%$,
consistent with the ethanol case.

\section{Effective Temperature of the MD17 Dataset}
\label{app:effective_temperature}

The MD17 dataset provides \textit{ab initio} molecular dynamics (AIMD)
trajectories computed at the DFT level of theory, and was generated at a nominal temperature of $T = 500\,\mathrm{K}$
\cite{Chmiela2017,Chmiela2018}. The data used in this work come from the
revised dataset rMD17~\cite{Christensen2020rMD17}, which keeps the MD17 geometries but
recomputes their energies and forces at the PBE/def2-SVP level. Our training
configurations are therefore snapshots of these nominally $500\,\mathrm{K}$
trajectories, and, as we show below, they are in fact distributed at a higher
temperature.
In the reduced units employed throughout this work, $T^{*} = 1$ corresponds
to the nominal $T = 500\,\mathrm{K}$, or equivalently to an inverse
temperature $\beta_{\mathrm{nom}} \approx 1.006\,\mathrm{mol\,kcal^{-1}}$.
The reduced temperature is then simply
$T^{*} = \beta_{\mathrm{nom}}/\beta_{\mathrm{eff}}$, so that an estimated $\beta_{\mathrm{eff}} < \beta_{\mathrm{nom}}$ directly signals an effective
temperature exceeding $500$\,$\mathrm{K}$.
In this appendix we show that the data are in fact hotter than this. We first describe how we noticed the problem, then explain it with the Nos\'e--Hoover thermostat used to generate the trajectories, and finally measure the effective temperature of ethanol and aspirin and check that sampling at this temperature fixes the problem.

\subsection*{The problem: the data are not stationary at the nominal temperature}

We first noticed the problem when MD and MCMC simulations of ethanol at the nominal $T^{*}=1$ did not reproduce the DFT energy distribution. To tell a wrong temperature apart from a wrong potential, we used a simple property of MCMC: a chain that satisfies detailed balance with respect to $p_\theta(\mathbf{x}) \propto e^{-\beta E_\theta(\mathbf{x})}$ leaves that distribution unchanged. So if we start the chains from reference configurations, and the model describes the energy surface well at the right temperature, the energy distribution should stay the same after each step. If the temperature is too low, the configurations drift towards lower energies; if it is too high, they drift towards higher energies.

\begin{figure}[htbp]
    \centering
    \includegraphics[width=\linewidth]{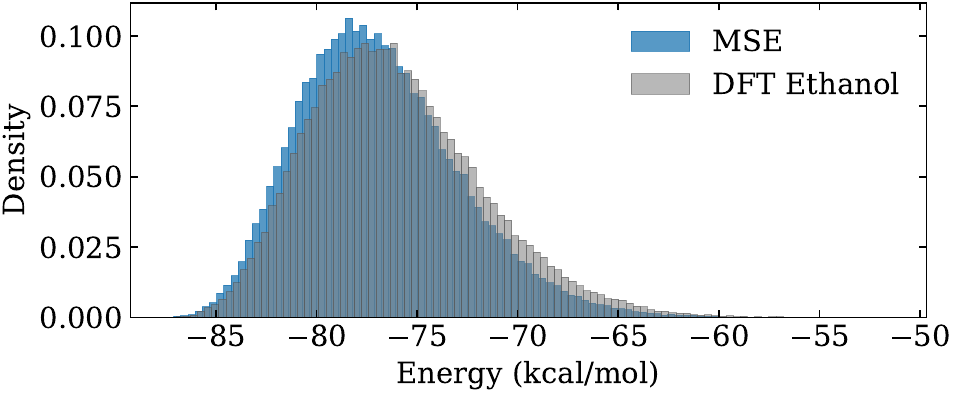}
    \caption{Energy distribution of $10^{4}$ ethanol
             configurations after a single MALA step under the MSE-trained
             potential at the nominal reduced temperature $T^{*}=1$ (blue),
             starting from DFT configurations, compared with the DFT reference
             (grey). The sampled distribution moves to lower energies, so $T^{*}=1$ is colder than the data.}
    \label{fig:stationarity_T1}
\end{figure}

We took $10^{4}$ DFT configurations of ethanol, and applied a single MALA step to each one with the MSE-trained potential. One step moves each configuration only slightly, so it cannot reach the spurious minima of Sec.~\ref{sec:results_failure} and stays close to the data, where the MSE potential is accurate (Table~\ref{tab:errors}). The test therefore checks the temperature and nothing else. At the nominal $T^{*}=1$ (Fig.~\ref{fig:stationarity_T1}), the energy distribution has already moved to lower energies after a single step. The configurations are cooling down, which means the data are hotter than the nominal temperature. Since all our sampling results depend on $\beta$, we had to understand and fix this before comparing anything with DFT.

\subsection*{The explanation: Nos\'e--Hoover thermostat and internal degrees of freedom}

Why would the data be hotter than $500\,\mathrm{K}$? The answer lies in the thermostat. The MD17 trajectories~\cite{Chmiela2017} were generated with a
Nos\'e--Hoover thermostat~\cite{Nose1984,Hoover1985} set to
$T_0=500\,\mathrm{K}$~\cite{Chmiela2018}.
For an isolated molecule, the equations of motion can be written as
\begin{equation}
    \dot{\mathbf R}_i=\frac{\mathbf p_i}{m_i},\qquad
    \dot{\mathbf p}_i=\mathbf F_i-\zeta\,\mathbf p_i,\qquad
    \dot\zeta=\frac{2K-g\,k_{\mathrm B}T_0}{Q}.
    \label{eq:nose_hoover}
\end{equation}
Here $\mathbf R_i$, $\mathbf p_i$, and $m_i$ are the position, momentum, and mass
of atom $i$; $\mathbf F_i=-\nabla_i U$ is the force derived from the molecular
potential; and $K=\sum_i\mathbf p_i^2/(2m_i)$ is the total kinetic energy. The
variable $\zeta$ describes the action of the thermostat, $Q$ controls its response
time, and $g$ is the number of degrees of freedom used to set its target kinetic
energy.

Let $\mathbf P=\sum_i\mathbf p_i$ and
$\mathbf L=\sum_i\mathbf R_i\times\mathbf p_i$ be the total linear and angular
momenta. Because the molecular potential is invariant under translation and
rotation, its forces exert neither a net force nor a net torque,
$\sum_i\mathbf F_i=\sum_i\mathbf R_i\times\mathbf F_i=0$. Since
$\sum_i\dot{\mathbf R}_i\times\mathbf p_i=0$, it follows from
Eq.~\eqref{eq:nose_hoover} that
\begin{equation}
    \dot{\mathbf P}=-\zeta\,\mathbf P,\qquad
    \dot{\mathbf L}=-\zeta\,\mathbf L.
\end{equation}
Thus, if the molecule starts with $\mathbf P=\mathbf L=0$, both remain zero: the
thermostat cannot generate either momentum from zero. Conservation laws of this
kind are known to change the distribution generated by Nos\'e--Hoover dynamics,
which is then no longer canonical at $T_0$ unless $g$ is chosen to account for
them~\cite{Tuckerman2001}. We find that both momenta are indeed negligible in the
MD17 trajectories of ethanol and aspirin. Their kinetic energy is therefore carried
almost entirely by the $3N-6$ internal degrees of freedom.

If the thermostat nevertheless counted all $g=3N$ Cartesian degrees
of freedom, it would target $\langle K\rangle\simeq(3N/2)\,k_{\mathrm B}T_0$. The
corresponding internal kinetic temperature would be
\begin{equation}
    T_{\mathrm{int}}
    =\frac{2\langle K\rangle}{(3N-6)\,k_{\mathrm B}}
    \simeq\frac{3N}{3N-6}\,T_0.
    \label{eq:T_int}
\end{equation}
This predicts approximately $643\,\mathrm{K}$ ($T^{*}=1.29$) for ethanol ($N=9$)
and $553\,\mathrm{K}$ ($T^{*}=1.11$) for aspirin ($N=21$). These predictions agree
closely with the temperatures obtained by reconstructing the kinetic energy from
consecutive configurations of the original MD17 trajectories: approximately
$649\,\mathrm{K}$ and $555\,\mathrm{K}$, respectively. To obtain these values, we
estimated the atomic velocities by forward finite differences between consecutive
frames, $\mathbf v_i(t)\approx[\mathbf R_i(t+\Delta t)-\mathbf R_i(t)]/\Delta t$ with
the $\Delta t=0.5\,\mathrm{fs}$ frame spacing of the trajectories~\cite{Chmiela2018},
computed the kinetic energy $K$, and applied Eq.~\eqref{eq:T_int}. The thermostat therefore explains both why the data are hotter than $500\,\mathrm{K}$ and by how much.

\subsection*{The solution: measuring the effective temperature}

Equation~\eqref{eq:T_int} gives the internal \emph{kinetic} temperature. Our sampler only sees configurations, so what matters for it is the \emph{configurational} temperature, which sets the distribution of the potential energy. In equilibrium the two are equal. As a quick and independent check, we estimate the configurational temperature directly from the DFT energies with the fluctuation--dissipation theorem (FDT).

\subsubsection*{Fluctuation–dissipation estimator}

This estimate uses the energy fluctuation–dissipation relation.
For a system with $N_{\mathrm{df}}$ vibrational degrees of freedom (i.e.\
$3N - 6$ after removing translational and rotational modes), the canonical
variance of the potential energy satisfies
\begin{equation}
    \langle (\delta E)^{2} \rangle
    = \frac{N_{\mathrm{df}}}{2\,\beta_{\mathrm{eff}}^{2}},
    \label{eq:fdt}
\end{equation}
which gives the estimator
\begin{equation}
    \beta_{\mathrm{FDT}} = \sqrt{\frac{N_{\mathrm{df}}}{2\,\mathrm{Var}(E)}}.
    \label{eq:beta_fdt}
\end{equation}
Equation~\eqref{eq:fdt} assumes that the internal modes are approximately harmonic. Applying Eq.~\eqref{eq:beta_fdt} to the DFT trajectory of each molecule
yields the FDT-based effective temperatures reported in
Table~\ref{tab:Teff}.

The FDT estimate agrees with the thermostat prediction of Eq.~\eqref{eq:T_int}. For ethanol, all the estimates are within $4\%$ of each other. For aspirin, the FDT value is about $5\%$ higher than the kinetic one. FDT assumes harmonic internal modes [Eq.~\eqref{eq:fdt}], and aspirin, being larger and more anharmonic, is likely to deviate more from this.
A second, smaller source of difference is that the configurations were sampled on the
PBE+vdW-TS surface of the original simulations~\cite{Chmiela2018}, whereas the energies we use are the rMD17 PBE/def2-SVP values~\cite{Christensen2020rMD17}. The close agreement
between the kinetic and configurational estimates suggests that this effect is small.

\subsection*{Validation: stationarity at the effective temperature}

\begin{figure}[htbp]
    \centering
    \includegraphics[width=\linewidth]{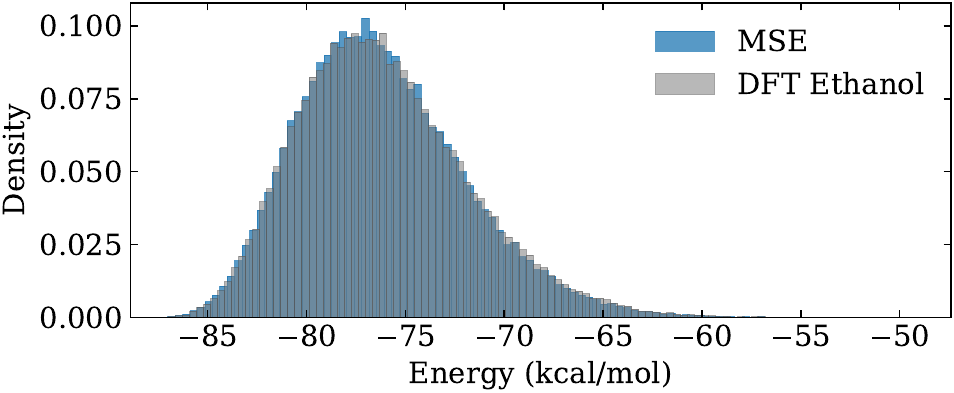}
    \caption{Same as Fig.~\ref{fig:stationarity_T1}, but at the effective temperature $T^{*}=1.3$. The two distributions now match, so the sampler leaves the data distribution unchanged at this temperature.}
    \label{fig:stationarity_T13}
\end{figure}

Finally, we go back to the test that revealed the problem and repeat it at $T^{*}=1.3$. The sampled and reference energy distributions now match (Fig.~\ref{fig:stationarity_T13}), so the sampler leaves the data distribution unchanged. The mismatch at $T^{*}=1$ was caused by the temperature, not by the potential. For this reason, all sampling in this work uses the effective temperature,
$k_BT = 1.3\,\mathrm{kcal\,mol^{-1}}$ ($T\approx650\,\mathrm{K}$) for ethanol and $k_BT = 1.155\,\mathrm{kcal\,mol^{-1}}$ ($T\approx580\,\mathrm{K}$) for aspirin, which lies within the range of our estimates ($T^{*} = 1.11$--$1.16$).

\subsection*{Summary}

Table~\ref{tab:Teff} collects all estimates alongside the nominal temperature.
All methods consistently place the effective temperature well above
$500\,\mathrm{K}$, reaching values close to $T^{*}\approx 1.3$ in reduced
units for ethanol and $T^{*}\approx 1.1$--$1.16$ for aspirin.
The difference comes from the way the thermostat counts degrees of freedom [Eq.~\eqref{eq:T_int}]: it sets the kinetic energy for $3N$ degrees of freedom, but this energy is shared among only the $3N-6$ internal ones.
More generally, because the same thermostat protocol underlies the
whole dataset, the effective temperature must be accounted for when
benchmarking models trained on MD17 against physical observables at
$500\,\mathrm{K}$.

\begin{table}[htbp]
    \centering
    \caption{Effective temperatures of the MD17 molecules studied in this
             work, estimated from the degree-of-freedom correction of Eq.~\eqref{eq:T_int}, from the kinetic energy reconstructed from consecutive configurations, and from the fluctuation--dissipation theorem (FDT). The nominal temperature reported in the literature is
             $T_{\mathrm{nom}} = 500\,\mathrm{K}$ ($T^{*}=1$).}
    \label{tab:Teff}
    \setlength{\tabcolsep}{6pt}
    \begin{tabular}{lcccc}
        \toprule
        & \multicolumn{2}{c}{Ethanol} & \multicolumn{2}{c}{Aspirin} \\
        Estimator & $T$ (K) & $T^{*}$ & $T$ (K) & $T^{*}$ \\
        \midrule
        Nominal            & 500 & 1.00 & 500 & 1.00 \\
        FDT                & 667 & 1.33 & 581 & 1.16 \\
        Eq.~\eqref{eq:T_int} & 643 & 1.29 & 553 & 1.11 \\
        Kinetic              & 649 & 1.30 & 555 & 1.11 \\
        \bottomrule
    \end{tabular}
\end{table}

\section{CRMSE Training for Aspirin}
\label{app:aspirin}
\begin{figure}[H]
    \centering
    \includegraphics[width=0.8\linewidth]{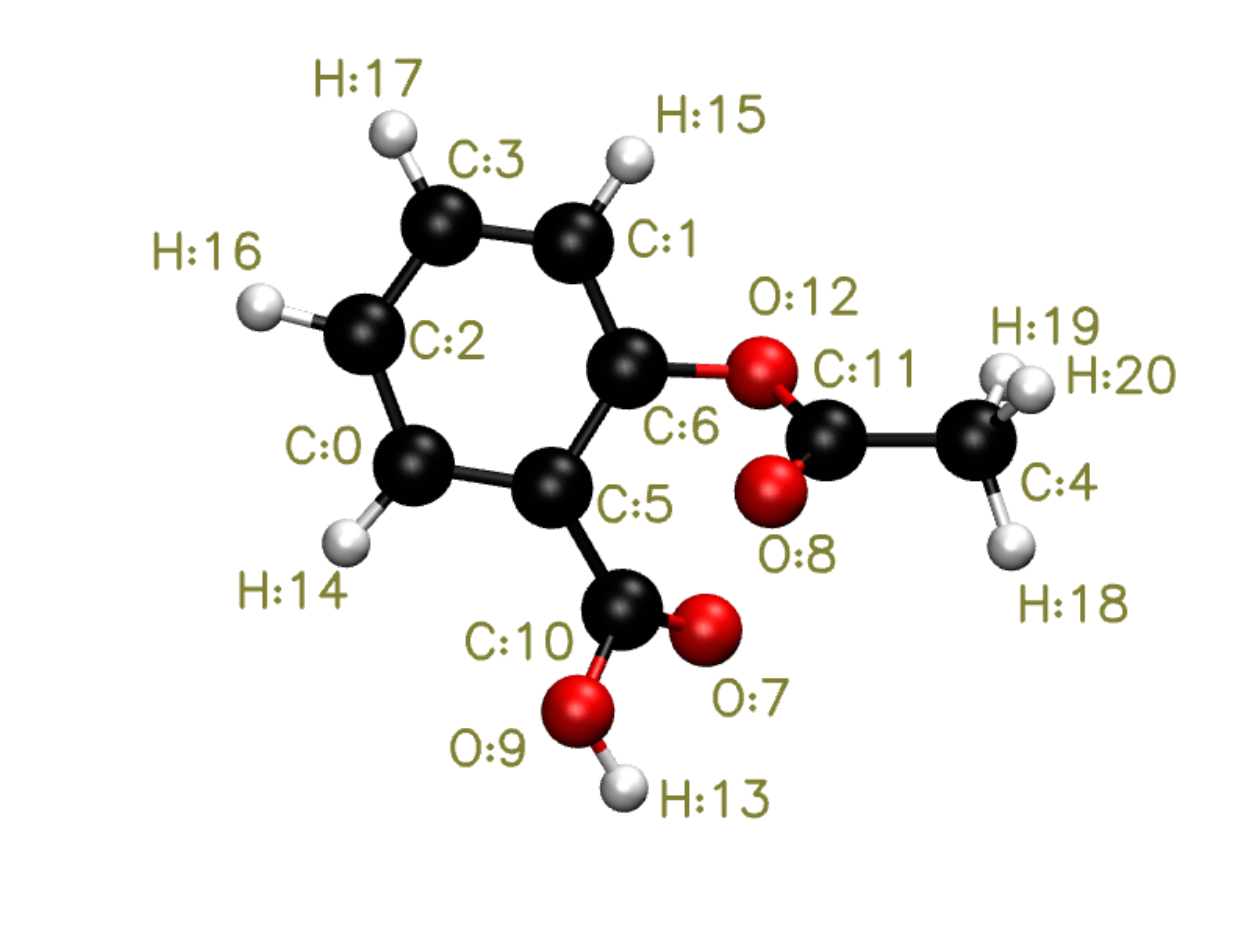}
    \caption{3D representation of the DFT aspirin molecule.}
    \label{fig:dft_aspirin}
\end{figure}

In this appendix, we implement the new regularization training for the aspirin molecule to demonstrate that the approach generalizes to other molecules. We follow the same steps as for the ethanol molecule, again training with the MSE using 950 training data points for aspirin and $50$ for validation (as described in the GNN-LF paper~\cite{pmlr-v198-wang22d}). In Fig.~\ref{fig:dft_aspirin} we show a DFT configuration of aspirin with the index and name of each atom. Note that the bond between
C$_{11}$ and O$_{8}$ is a double bond, which is not rendered
explicitly in this figure or in any of the subsequent molecular visualizations of aspirin.

\begin{figure}[h!]
\centering
\captionsetup{justification=raggedright, singlelinecheck=false, format=plain}
    \includegraphics[width=1.\linewidth]{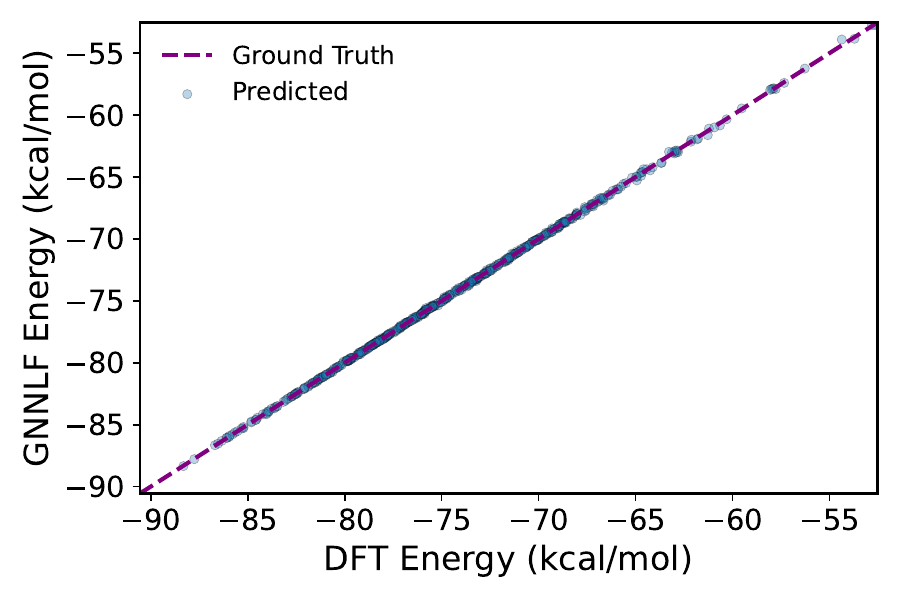}
    \includegraphics[width=1.\linewidth]{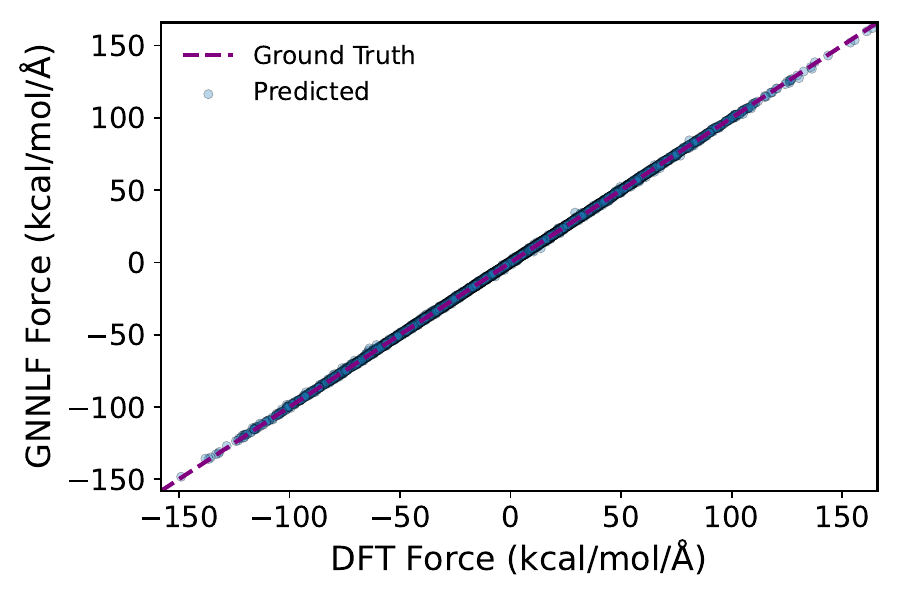}
    \caption{Performance of the MSE-trained GNN-LF potential on 1000 held-out DFT configurations: predicted versus reference energies (top) and forces (bottom).}
    \label{fig:mse_perf_aspirin}
\end{figure}

As in the previous case, the energy and force predictions show strong agreement with the ideal line, as illustrated in Fig.~\ref{fig:mse_perf_aspirin}.

\begin{figure}[H]
    \centering
    \includegraphics[width=0.8\linewidth]{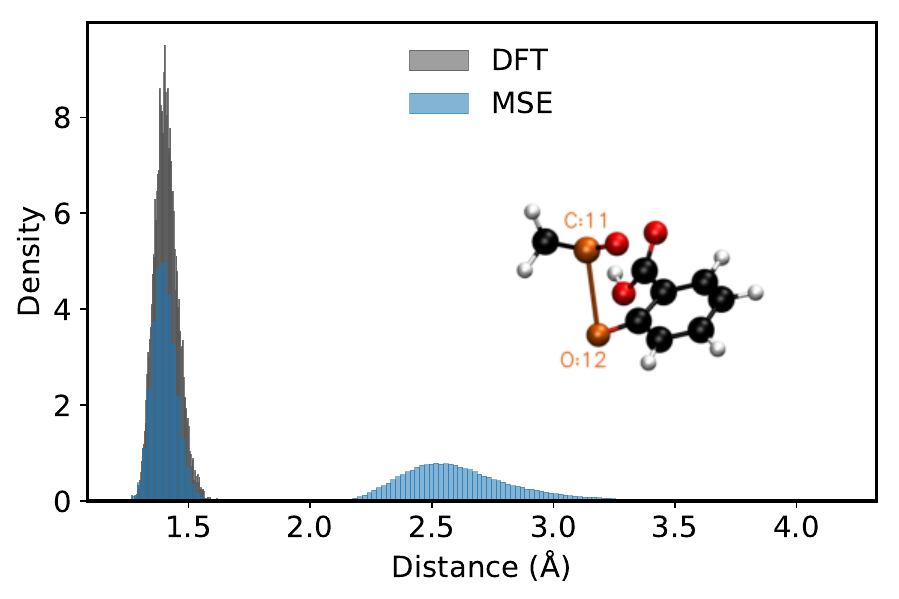}
    \caption{Density of interatomic distances for selected atom pairs under MSE-MALA sampling (blue) and the DFT reference (gray).}
    \label{fig:mse_obs_aspirin}
\end{figure}

However, this good performance on the physical quantities does not survive sampling. We incorporate the learned potential into MALA at the inverse temperature $\beta$ corresponding to $k_BT = 1.155$~kcal/mol ($T^{*}=1.155$, $T \approx 580$~K), the effective temperature at which the MD17 reference trajectory was generated, using a global update with a per-atom preconditioning matrix (defined in Appendix~\ref{app:precond}). Initializing each chain from a reference DFT configuration, we run 32 independent simulations of $10^6$ steps with step size $0.0006$. Despite the model's predictive fidelity, the trajectories escape the physical basin and settle into spurious low-energy minima absent from the reference data; the 32 chains are mutually consistent, so this reflects a systematic failure of the potential rather than chain-to-chain variance. In Fig.~\ref{fig:mse_obs_aspirin} we show that the resulting observable of distances between some specific atoms diverges, leading to unphysical distances and therefore unrealistic representations of the aspirin molecule.

\begin{figure}[h!]
    \centering
    \captionsetup{justification=raggedright, singlelinecheck=false, format=plain}
    \includegraphics[width=1.\linewidth]{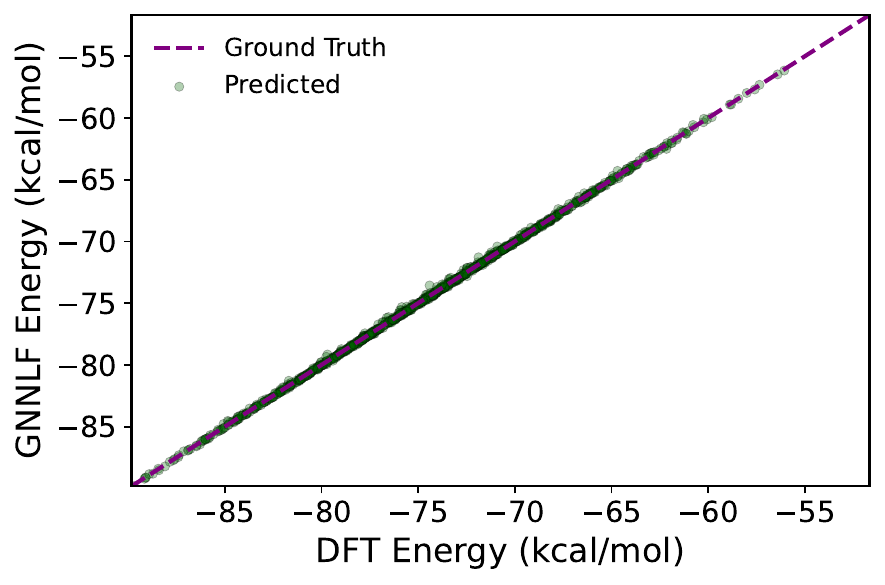}
    \includegraphics[width=1.\linewidth]{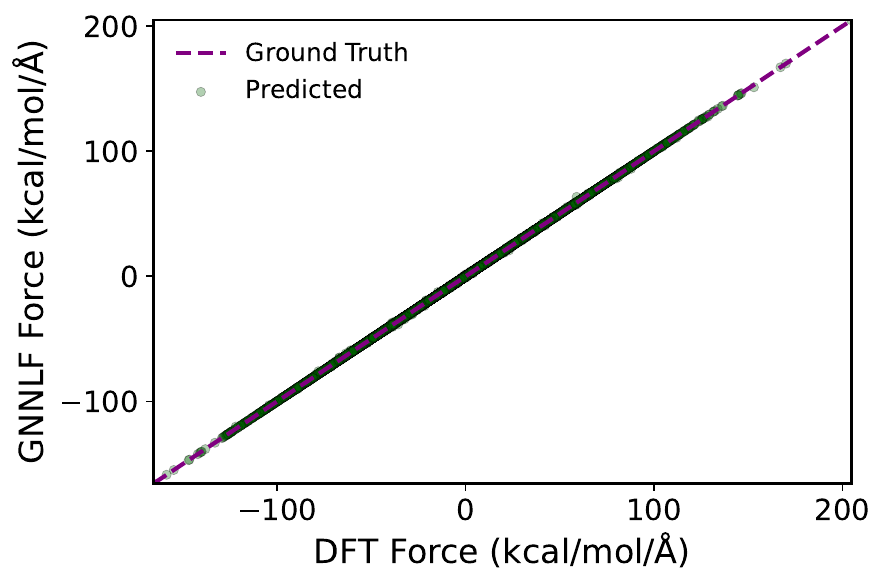}
    \caption{Performance of the CRMSE post-trained model on the held-out DFT test set: predicted versus reference energies (top) and forces (bottom). The force accuracy is essentially unchanged from that of the MSE-pretrained model (Fig.~\ref{fig:mse_perf_aspirin}), and the energies remain close to the reference (Table~\ref{tab:errors-aspirin}).}
    \label{fig:crmse_perf_aspirin}
\end{figure}

To address this issue, we continue the training of the surrogate model using the same data, but now implementing the new Contrastive Regularization (CR). We collect out-of-distribution configurations from an MSE-MALA chain and insert them into the replay buffer, which has a size of 1000. We then restart the training with Adam, restoring the optimizer state (moments and learning rate, ${\approx}10^{-5}$) from the last epoch of the MSE training. We train with the new regularizer until convergence, up to epoch 200, where the CR regularizer has a strength of $0.015$. We use the ULA for the CRMSE procedure with the same preconditioning matrix and a step size prefactor of $0.0004$. We also reinitialize each chain drawn from $\mathcal{B}$, with probability
$\rho$, to a random training configuration before its $K$ ULA steps, similar to the random restarts used with persistent replay buffers~\cite{Du2019ImplicitGA}, but starting from data as in contrastive divergence~\cite{Hinton2002CD}.
The reinitialization rate $\rho$ can be tuned: for aspirin, $\rho = 0.02$
lowers the held-out energy MAE from $0.43$ to $0.10$~kcal/mol relative to
pure PCD ($\rho = 0$), with the force MAE unchanged. All ethanol results use $\rho = 0$. The energy and force predictions of the CRMSE post-trained model are reported
in Fig.~\ref{fig:crmse_perf_aspirin}: both the energies and the forces remain tightly aligned with the reference line.
The corresponding MAEs are reported in Table~\ref{tab:errors-aspirin}.
\begin{table}[h!]
    \centering
    \caption{Energy and force errors of the MSE-pretrained and CRMSE post-trained models on the 1000-configuration held-out DFT test set. The force errors are essentially unchanged, and the energy error rises only slightly ($0.06$ to $0.10$~kcal/mol), well within chemical accuracy.}
    \label{tab:errors-aspirin}
    \begin{ruledtabular}
    \begin{tabular}{lcc}
         & Energy MAE & Force MAE \\
         & (kcal/mol) & (kcal/mol/\AA) \\
        \colrule
        MSE   & 0.06 & 0.20 \\
        CRMSE & 0.10 & 0.21 \\
    \end{tabular}
    \end{ruledtabular}
\end{table}

\begin{figure}[H]
    \centering
    \includegraphics[width=0.8\linewidth]{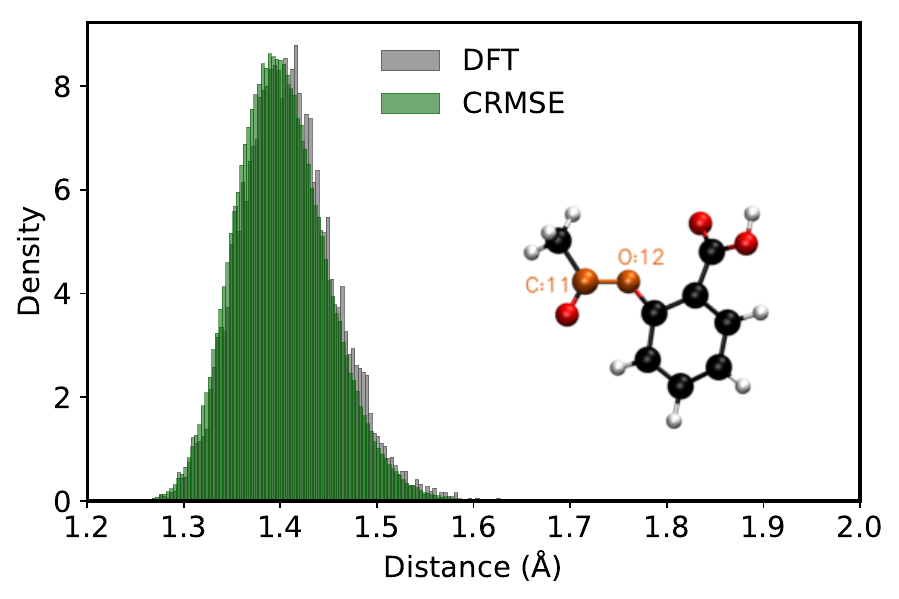}
    \caption{Density of interatomic distances for selected atom pair under CRMSE-MALA sampling (green) and the DFT reference (gray).}
    \label{fig:crmse_obs_aspirin}
\end{figure}

After training with the new model, we perform experiments incorporating the trained model into the MALA. We perform 32 independent simulations for $10^6$ global steps and we illustrate the results again by depicting the density of interatomic distances for the same selected atom-pair, which can be seen in Fig.~\ref{fig:crmse_obs_aspirin}.

\section{CRMSE training for a smaller training dataset}
\label{app:appendixE}
\begin{figure}[h!]

\centering

\captionsetup{justification=raggedright, singlelinecheck=false, format=plain}

\includegraphics[width=1.\linewidth]{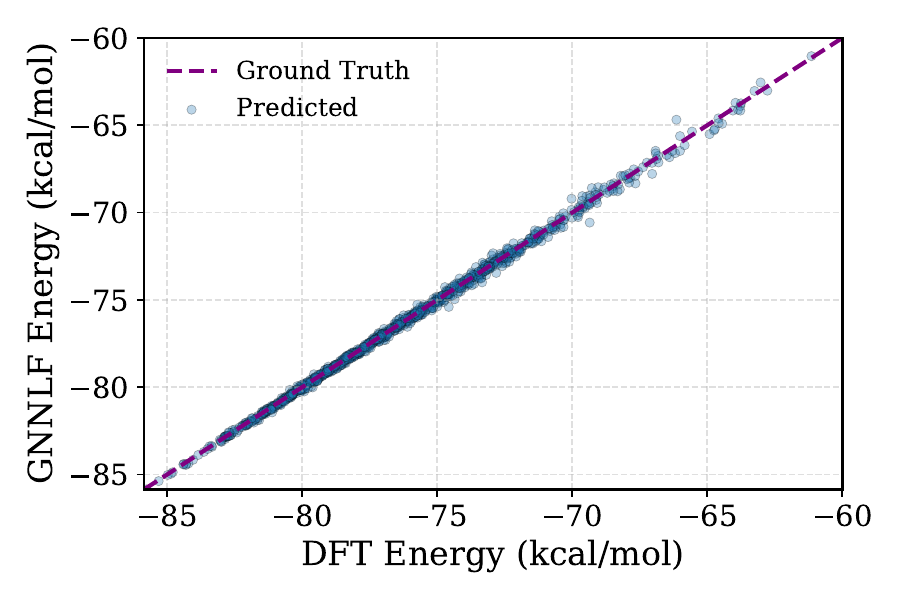}

\includegraphics[width=1.\linewidth]{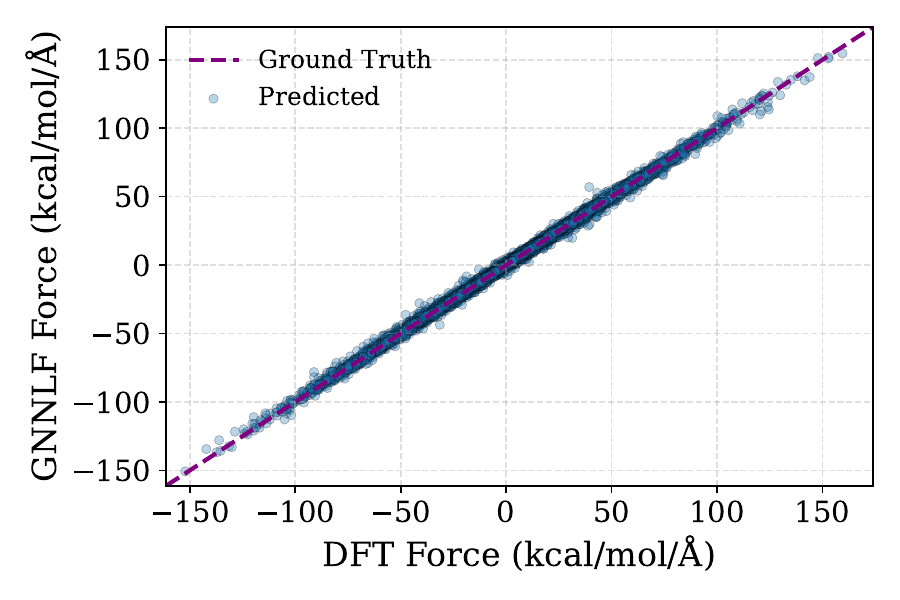}

\caption{Performance of the MSE-trained GNN-LF potential (trained on 200 configurations) on 1000 held-out DFT configurations: predicted versus reference energies (top) and forces (bottom).}

\label{fig:mse_perf_200dt}

\end{figure}

In this Appendix, we demonstrate the CRMSE training procedure using a reduced dataset, following the methodology outlined in the Results section. Starting from a training set of 200 DFT configurations, the MSE baseline model achieves reasonable performance on both energy and forces, albeit with a larger MAE compared to the full-dataset case, as shown in Fig.~\ref{fig:mse_perf_200dt}.

\begin{figure}[h!]

\centering

\captionsetup{justification=raggedright, singlelinecheck=false, format=plain}

\includegraphics[width=1.\linewidth]{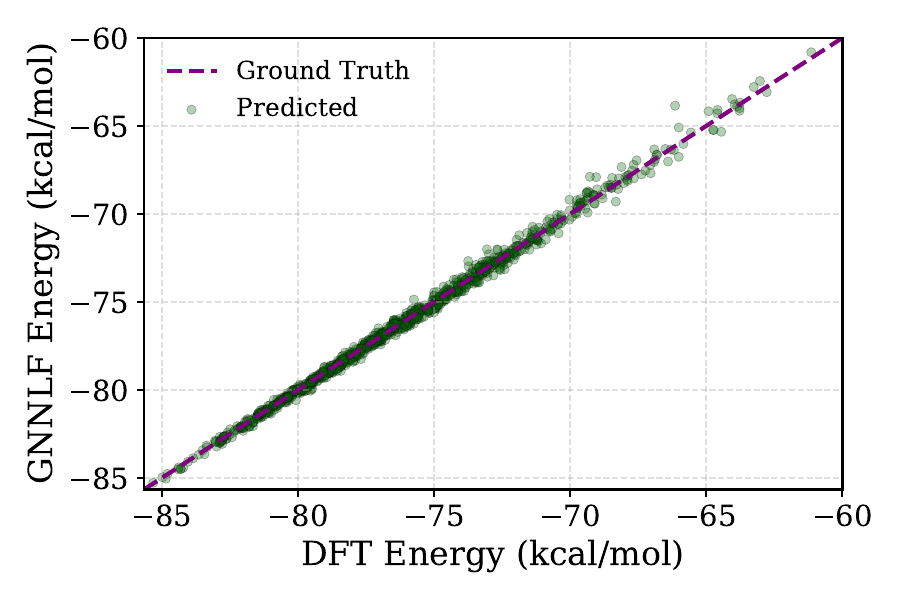}

\includegraphics[width=1.\linewidth]{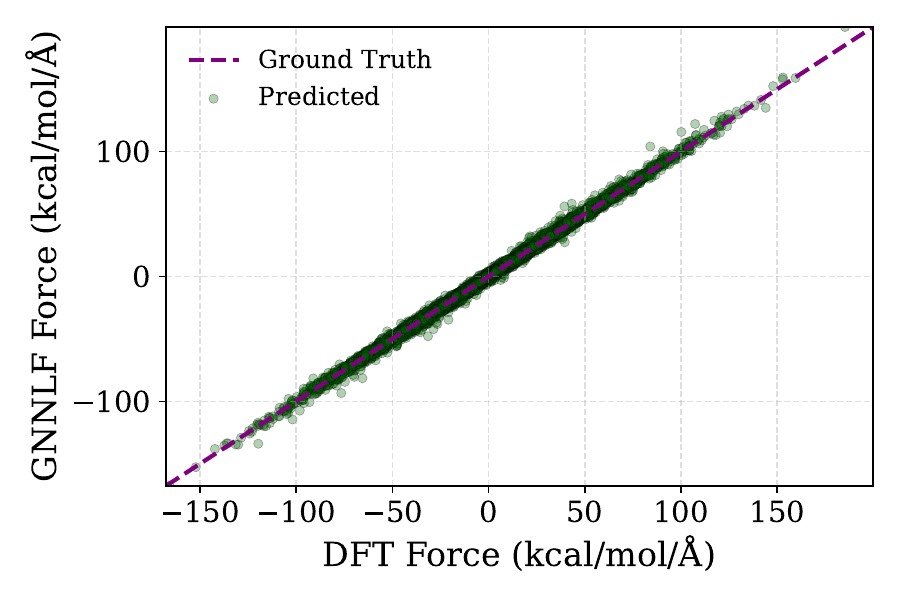}

\caption{Performance of the CRMSE-trained GNN-LF potential (trained on 200 configurations) on 1000 held-out DFT configurations: predicted versus reference energies (top) and forces (bottom).}

\label{fig:crmse_perf_200}

\end{figure}

We then post-train this model using CRMSE, employing a replay buffer of 1000
configurations sampled from MALA simulations run with the MSE model. Training is
carried out for 500 epochs with a regularization parameter $\lambda = 0.05$ (within the stable window $[0.01, 0.5]$), $K = 200$
Langevin steps, and a ULA step size of $0.0002$. The energy and force predictions of the resulting CRMSE model are reported in Fig.~\ref{fig:crmse_perf_200},
with the corresponding MAE summarized in Table~\ref{tab:errors-200data}, showing performance close to that of the full-dataset MSE model.

\begin{table}[h!]
    \centering
    \caption{Energy and force errors of the 200 data MSE-pretrained and CRMSE post-trained models on the 1000-configuration held-out DFT test set. Both models predict energies and forces within chemical accuracy, confirming that the contrastive regularization preserves predictive accuracy on the data-constrained region.}
    \label{tab:errors-200data}
    \begin{ruledtabular}
    \begin{tabular}{lcc}
         & Energy MAE & Force MAE \\
         & (kcal/mol) & (kcal/mol/\AA) \\
        \colrule
        MSE   & 0.13 & 0.71 \\
        CRMSE & 0.19 & 0.93 \\
    \end{tabular}
    \end{ruledtabular}
\end{table}

\section{Stiffness selection and chain-validity decomposition for the geometric prior}
\label{app:prior_decomposition}

This appendix explains how we chose the stiffness $(k_d,k_\alpha)$ of the geometric prior of Sec.~\ref{sec:results_prior_reweighting}, and splits the resulting free-energy profile according to whether each chain stayed within the prior's bounds.

\subsection{Stiffness scan}
\label{app:prior_stiffness_scan}

We scanned six $(k_d,k_\alpha)$ settings covering the usable stiffness range. For each one, we added the prior of Eq.~\eqref{eq:prior} to the MSE-trained potential, ran 32 independent MALA chains, and collected the diagnostics in Table~\ref{tab:prior_stiffness_ablation_full}. With soft walls, the chains move further outside the training range, so they hit the walls more often and the sampled bond-length and angle distributions drift further from the reference. With very stiff walls, wall contacts and chain-to-chain variance go up again, probably because the prior gets close to the stability limit of the integrator instead of confining the molecule better. The setting $(k_d,k_\alpha)=(5000,5000)$ gives the lowest bias and the fewest wall contacts on every metric in Table~\ref{tab:prior_stiffness_ablation_full}, and we use it in Sec.~\ref{sec:results_prior_reweighting} and in the decomposition below.

\begin{table*}[t]
\centering
\caption{Effect of the prior stiffness on the remaining distributional bias and on wall contacts, over 32 independent MALA chains, for six bond/angle stiffness settings. KS and Wasserstein ($W_1$) values are averaged over all bond or angle coordinates; max KS, max bond contact and max angle contact are the largest single per-coordinate values for each setting. The mean prior energy is averaged over all steps and chains. The best value in each column is in bold.}
\label{tab:prior_stiffness_ablation_full}
\begin{adjustbox}{width=\textwidth}
\begin{tabular}{c c c c c c c c c c}
\toprule
$k_{d}$ & $k_{\alpha}$ & Mean prior energy & Mean bond KS & Mean bond $W_1$ & Mean angle KS & Mean angle $W_1$ & Max KS & Max bond contact & Max angle contact \\
\midrule
2000  & 5000  & $0.0022 \pm 0.0120$ & 0.077 & 0.0106 & 0.090 & 0.0422 & 0.156 & 0.087\% & 0.097\% \\
5000  & 2000  & $0.0038 \pm 0.0205$ & 0.079 & 0.0112 & 0.098 & 0.0446 & 0.161 & 0.116\% & 0.270\% \\
5000  & 5000  & $\mathbf{0.0001 \pm 0.0003}$ & $\mathbf{0.075}$ & $\mathbf{0.0100}$ & $\mathbf{0.075}$ & $\mathbf{0.0362}$ & $\mathbf{0.127}$ & $\mathbf{0.005\%}$ & $\mathbf{0.006\%}$ \\
10000 & 10000 & $0.0012 \pm 0.0068$ & 0.079 & 0.0111 & 0.087 & 0.0411 & 0.158 & 0.194\% & 0.328\% \\
10000 & 5000  & $0.0055 \pm 0.0305$ & 0.078 & 0.0113 & 0.086 & 0.0427 & 0.158 & 0.466\% & 1.041\% \\
5000  & 10000 & $0.0012 \pm 0.0066$ & 0.080 & 0.0114 & 0.098 & 0.0442 & 0.172 & 0.118\% & 0.149\% \\
\bottomrule
\end{tabular}
\end{adjustbox}
\end{table*}

\subsection{Decomposition by chain validity}
\label{app:prior_decomp_detail}

\begin{figure}[h]
    \centering
    \includegraphics[width=1\linewidth]{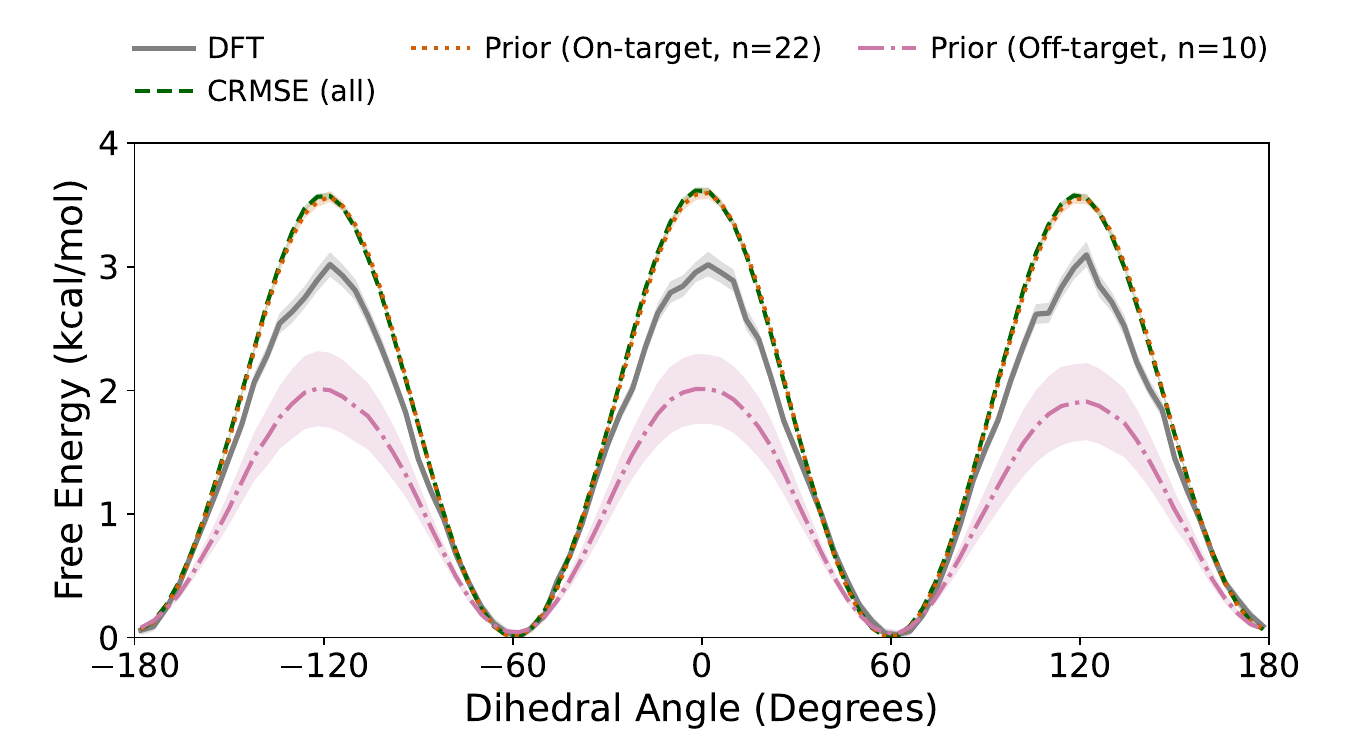}
    \caption{Free-energy profile along the $\mathrm{H\!-\!C\!-\!C\!-\!O}$ dihedral for the MSE potential with the geometric prior at the best stiffness $(k_d,k_\alpha)=(5000,5000)$ (Table~\ref{tab:prior_stiffness_ablation_full}), averaged over MALA chains and split by chain behavior. Chains that never crossed the prior's bounds (``on-target'', $n=22$), which are effectively plain MSE runs, give a barrier close to the CRMSE one instead of the low MSE barrier. Chains that crossed the bounds at least once (``off-target'', $n=10$) give a flat, underestimated profile close to that of the uncorrected MSE potential.}
    \label{fig:prior_decomposition}
\end{figure}

With this setting, we split the 32 chains into two groups: chains whose internal coordinates never left the prior's bounds (``on-target'', $n=22$), and chains that crossed the bounds at least once (``off-target'', $n=10$). Fig.~\ref{fig:prior_decomposition} shows the chain-averaged $\mathrm{H\!-\!C\!-\!C\!-\!O}$ profile for each group, together with the CRMSE and DFT references.

For the on-target chains the prior is never active, so it adds no energy or force at any step: these chains are ordinary MSE-MALA runs. Even so, their profile is close to the slightly overestimated CRMSE barrier, and not to the underestimated MSE barrier. We think this reflects a general trend. Once the configurations behind the MSE potential's spurious low-barrier pathway are removed from the sampled ensemble (here simply because these chains never reach them), the remaining trajectories tend to overestimate the barrier slightly instead of reproducing DFT exactly. The off-target chains give a flat, underestimated profile close to that of the uncorrected MSE potential, as expected once the local bond geometry becomes unphysical and the torsion loses its meaning.

CRMSE's own profile also overestimates the barrier slightly, with the same sign (Secs.~\ref{sec:results_freeenergy} and~\ref{sec:ula_validation}), and by about the same amount as the on-target chains. This is what we would expect: in both cases, the configurations responsible for the spurious low-barrier pathway are out of the sampler's reach. CRMSE achieves this by raising their energy far above the physical basin (Sec.~\ref{sec:results_mechanism}), while the on-target chains simply never enter that region. The two cases still differ in one important way. CRMSE keeps all 32 chains in the physical basin (Sec.~\ref{sec:results_crmse}), so it has no off-target chains. With the prior, a comparable barrier comes only from the 22 chains that happen never to touch the walls, while the other 10 end up in the flat regime of Fig.~\ref{fig:prior_decomposition}. The prior therefore works for some chains only, and CRMSE works for all of them.

\section{Free-energy profile along the hydroxyl (O--H) rotor}
\label{app:oh_rotor}

In the main text, we test the ethanol thermodynamics mainly through one torsion, the $\mathrm{H\!-\!C\!-\!C\!-\!O}$ methyl rotor. Ethanol has a second, independent internal rotation, the hydroxyl rotor: the $\mathrm{C\!-\!C\!-\!O\!-\!H}$ dihedral $\chi$, which describes the rotation of the O--H group about the C--O bond. Here we report its free-energy profile $F(\chi) = -k_B T \ln p(\chi)$ for every ethanol setting used in this work: MALA with the full training set
(Sec.~\ref{sec:results_freeenergy}), MALA with the reduced $200$-configuration set (Sec.~\ref{sec:thermo_200}), and BAOAB Langevin dynamics
(Sec.~\ref{sec:ula_validation}).

\subsection{The hydroxyl torsion of ethanol}
\label{app:oh_physics}

Rotation about the C--O bond of ethanol gives a well-known profile with three wells. The O--H group can point \emph{anti} (trans), near $\chi \approx 180^\circ$, or in one of two equivalent \emph{gauche} positions near $\chi \approx \pm 60^\circ$. The wells are separated by barriers of about $1$~kcal/mol, the highest one at the syn (eclipsed) geometry $\chi \approx 0^\circ$~\cite{kakar1980ethanol,pearson1997ethanol}. These are potential-energy barriers; the free-energy barriers in our profiles at $T \approx 650$~K are somewhat higher (about $1.7$~kcal/mol at the syn geometry in the DFT reference, Fig.~\ref{fig:oh_mala}). The anti and gauche wells are close enough in energy that all three are populated at our temperatures, so the balance between conformers depends on the whole shape of $F(\chi)$ and not on a single barrier. A potential that flattens or distorts this profile will get the populations wrong even if it describes the stiffer methyl-rotor barrier well. This makes the hydroxyl rotor a demanding and largely independent test of the corrected landscape.

\subsection{MALA sampling, full training set}
\label{app:oh_mala_full}

\begin{figure}[h]
    \centering
    \begin{subfigure}{\linewidth}
        \centering
        \includegraphics[width=1\linewidth]{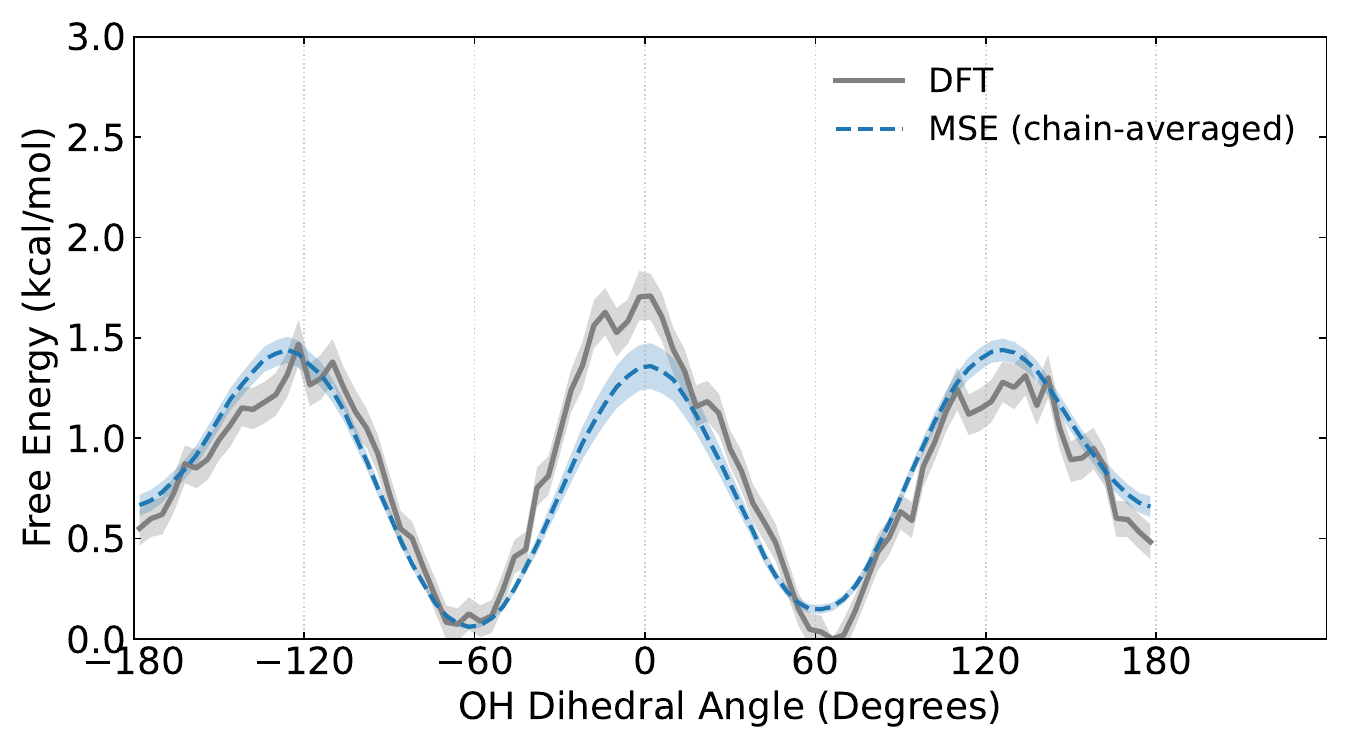}
        \caption{MSE-trained model}
        \label{fig:oh_mala_mse}
    \end{subfigure}

    \vspace{0.3cm}

    \begin{subfigure}{\linewidth}
        \centering
        \includegraphics[width=1\linewidth]{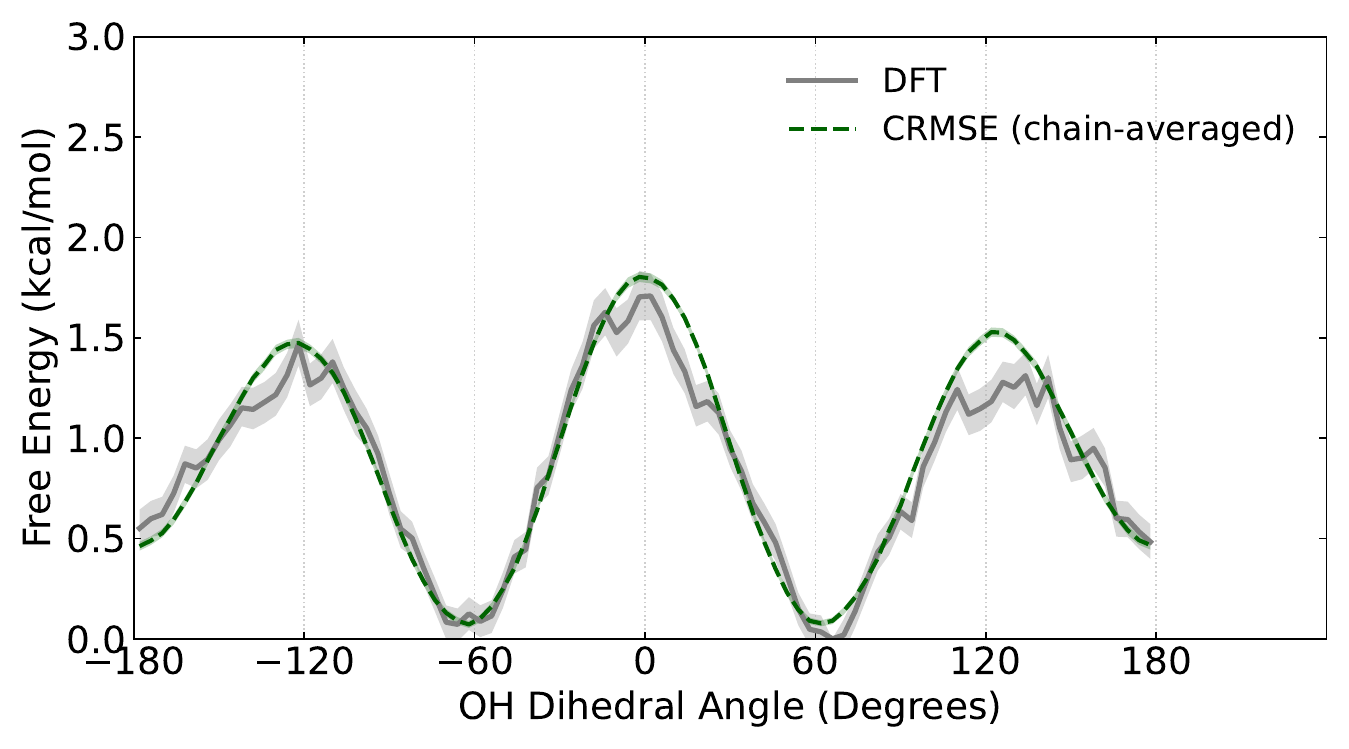}
        \caption{CRMSE post-trained model}
        \label{fig:oh_mala_crmse}
    \end{subfigure}

    \caption{Free-energy profile along the hydroxyl (O--H) rotor
    dihedral $\chi$ of ethanol under MALA sampling with the full training set,
    chain-averaged over the $32$ independent chains ($\pm 1$~SEM shaded). (a) The MSE model has the right overall shape but underestimates the barriers. (b) CRMSE brings the profile inside the DFT uncertainty band.}
    \label{fig:oh_mala}
\end{figure}

Fig.~\ref{fig:oh_mala} compares the chain-averaged MSE and CRMSE profiles $F(\chi)$ with the DFT reference, for the potentials trained on the full ethanol dataset and sampled with MALA (the same $32$ chains as in Sec.~\ref{sec:results_freeenergy}). Here the MSE potential already does reasonably well. It has the right three-well shape and the minima in the right places, but it underestimates the syn barrier ($\chi \approx 0^\circ$) by $\approx 0.3$~kcal/mol, and the shoulders at $\chi \approx \pm 120^\circ$ by a little less. CRMSE brings the whole profile inside the DFT uncertainty band, with only a slight ($\approx 0.1$~kcal/mol) overestimate of the central barrier, of the same sign as for the methyl rotor.

\subsection{MALA sampling, 200-configuration training set}
\label{app:oh_mala_200}

\begin{figure}[h]
    \centering
    \begin{subfigure}{\linewidth}
        \centering
        \includegraphics[width=1\linewidth]{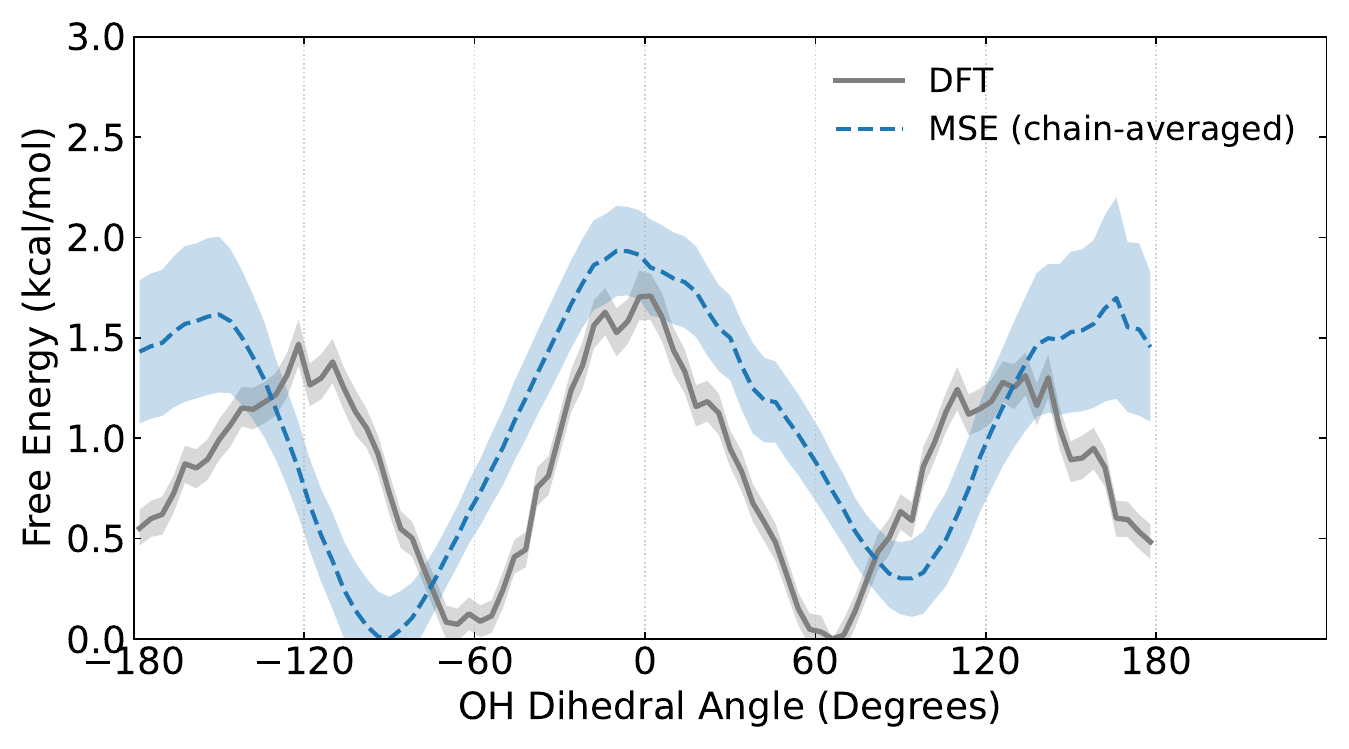}
        \caption{MSE-trained model}
        \label{fig:oh_mala_200_mse}
    \end{subfigure}

    \vspace{0.3cm}

    \begin{subfigure}{\linewidth}
        \centering
        \includegraphics[width=1\linewidth]{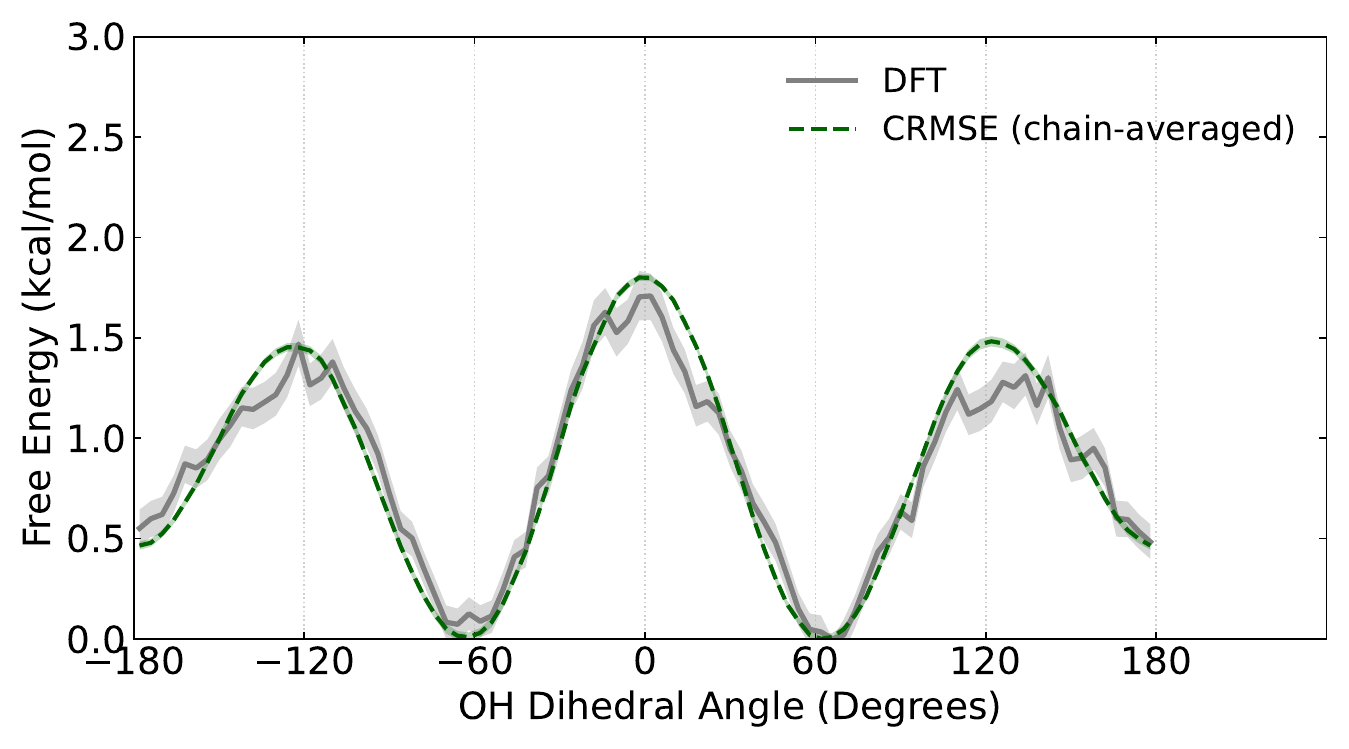}
        \caption{CRMSE post-trained model}
        \label{fig:oh_mala_200_crmse}
    \end{subfigure}

    \caption{Free-energy profile along the hydroxyl (O--H) rotor
    dihedral $\chi$ of ethanol under MALA sampling with the reduced
    $200$-configuration training set, chain-averaged over the $32$ independent
    chains ($\pm 1$~SEM shaded). (a) The MSE profile is strongly distorted and asymmetric, with a wide $\pm 1$~SEM band. (b) CRMSE gives almost the same profile as with the full data (Fig.~\ref{fig:oh_mala}).}
    \label{fig:oh_mala_200}
\end{figure}

Fig.~\ref{fig:oh_mala_200} repeats the comparison for the potentials trained on only $200$ DFT configurations (Sec.~\ref{sec:thermo_200}). With so little data the MSE result gets much worse: the profile becomes strongly asymmetric, its wells and barriers shift in angle, the syn barrier is overestimated, and the $\pm 1$~SEM band grows to $\approx \pm 0.5$~kcal/mol, so the chains no longer agree with each other. The CRMSE profile, on the other hand, is almost the same as with the full data (Fig.~\ref{fig:oh_mala}): barrier heights and well positions match DFT within its statistical uncertainty, and the spread between chains is back to the line width. The hydroxyl rotor is therefore as robust to reduced training data as the methyl rotor (Sec.~\ref{sec:thermo_200}).

\subsection{BAOAB Langevin dynamics}
\label{app:oh_baoab}

\begin{figure}[h]
    \centering
    \begin{subfigure}{\linewidth}
        \centering
        \includegraphics[width=1\linewidth]{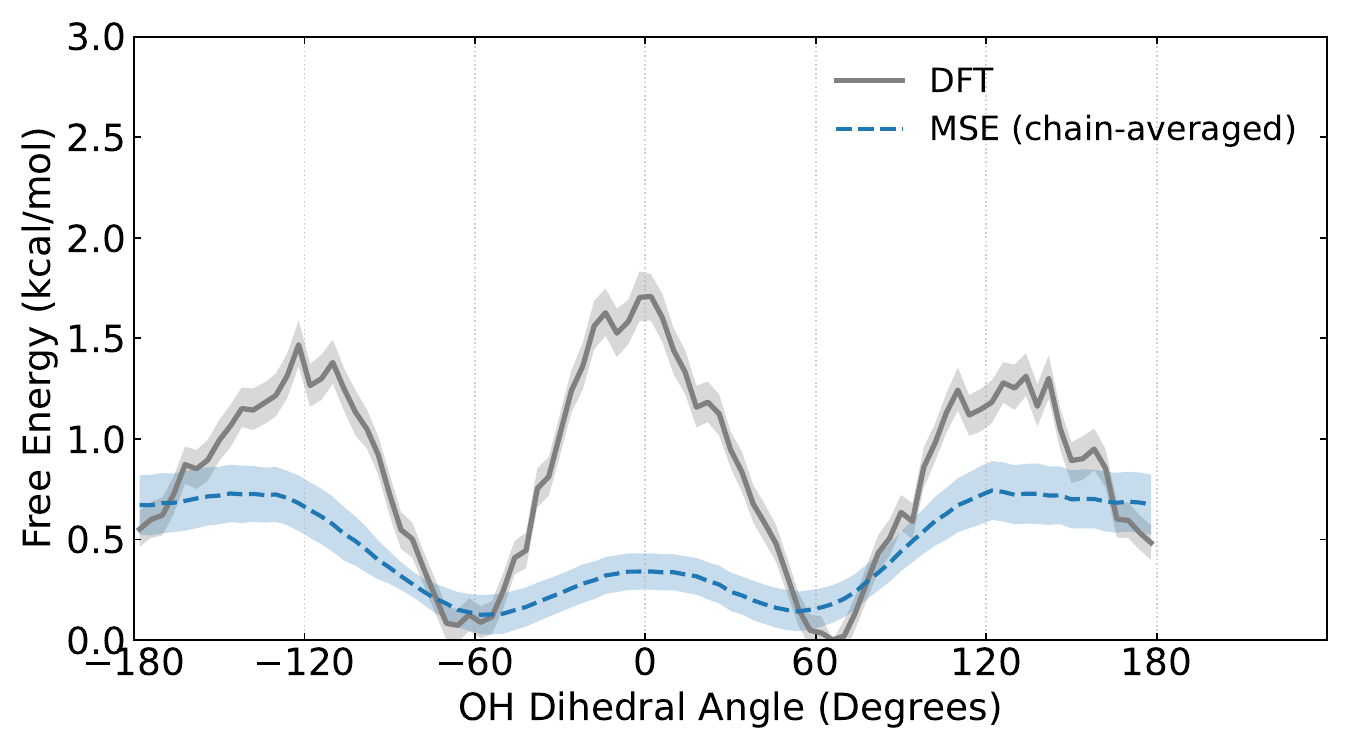}
        \caption{MSE-trained model}
        \label{fig:baoab_oh_mse}
    \end{subfigure}

    \vspace{0.3cm}

    \begin{subfigure}{\linewidth}
        \centering
        \includegraphics[width=1\linewidth]{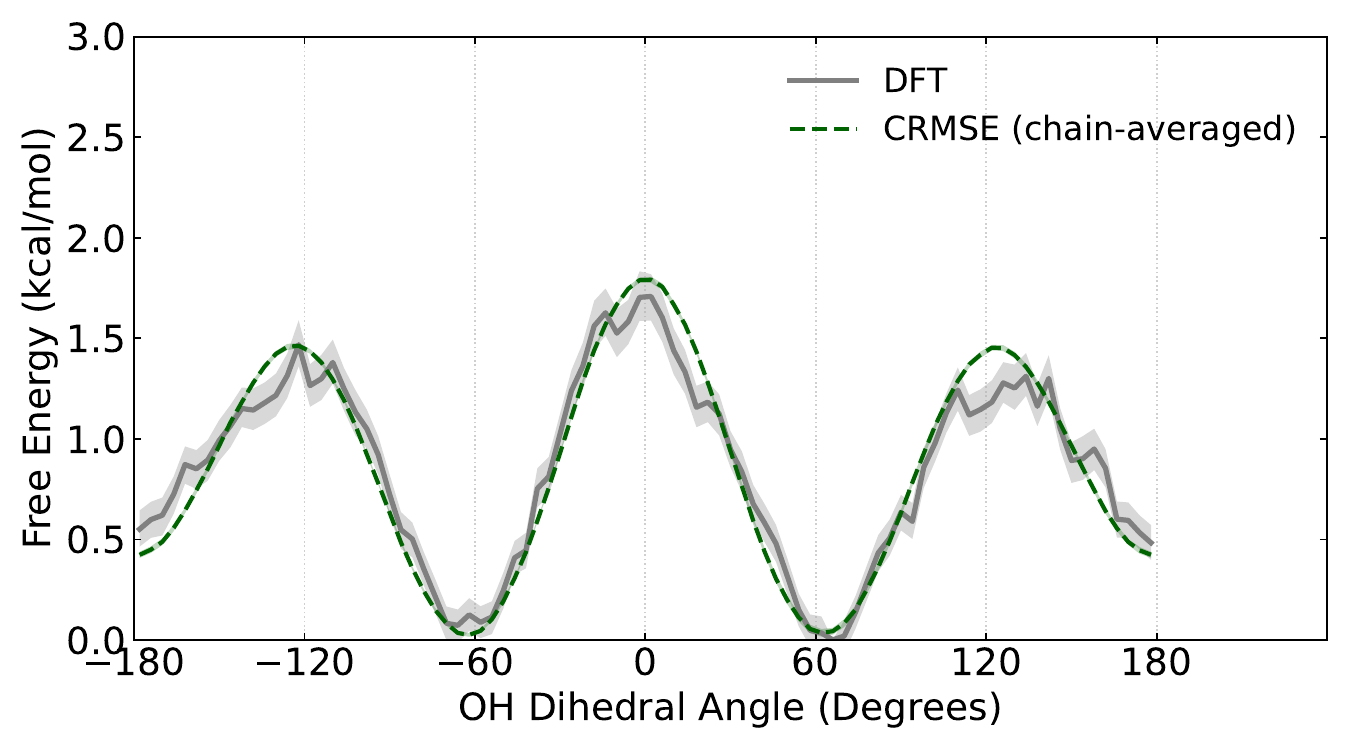}
        \caption{CRMSE post-trained model}
        \label{fig:baoab_oh_crmse}
    \end{subfigure}

    \caption{Free-energy profile along the hydroxyl (O--H) rotor
    dihedral $\chi$ of ethanol under BAOAB Langevin dynamics (same $10$ chains and $\pm1$~SEM shading as Fig.~\ref{fig:baoab_methyl}). (a) The MSE model flattens the profile far below the DFT reference and removes the barrier between the anti and gauche wells. (b) CRMSE restores the full profile.}
    \label{fig:baoab_oh}
\end{figure}

Fig.~\ref{fig:baoab_oh} shows the same comparison under BAOAB Langevin dynamics, using the $10$ independent $5$~ns chains of Sec.~\ref{sec:ula_validation}. Here the MSE failure is at its worst, most likely because these simulations have no accept/reject step and are much longer: each of the 10 BAOAB chains runs for $10^7$ steps, against $10^6$ steps for each MALA chain, which gives the sampler more time to reach the spurious regions of the MSE potential. The profile is flattened far below the DFT reference and the syn barrier drops from $\approx 1.7$ to $\approx 0.35$~kcal/mol, so the barrier between the anti and gauche wells almost disappears and the molecule no longer stays in one conformer. CRMSE restores the full three-well profile, and here it agrees with DFT better than for any other dihedral in this study.

In all three settings, the hydroxyl rotor therefore behaves like the methyl rotor, even though it was not used to tune the method. MSE moves away from the reference, slightly with full-data MALA and strongly with little data or under BAOAB, and CRMSE brings it back.

\clearpage
\bibliographystyle{apsrev4-2}
\bibliography{bibliography}    %

\end{document}